\documentclass[sigconf]{acmart}
\renewcommand\footnotetextcopyrightpermission[1]{}

\usepackage{comment}
\usepackage{pifont}
\usepackage{algorithmic}
\usepackage{adjustbox}
\usepackage{multirow}
\usepackage{tabularx}
\usepackage{enumitem}
 \usepackage{tcolorbox}
\usepackage{framed}
\usepackage[linesnumbered,ruled]{algorithm2e}
\SetKwInput{KwInput}{Input}          
\SetKwInput{KwOutput}{Output}  
\SetKwInput{KwProcedure}{Procedure} 
\SetCommentSty{mycommfont}
\usepackage[colorinlistoftodos]{todonotes}
\usepackage{graphicx}
\usepackage{textcomp}
\usepackage{xcolor}
\usepackage{multicol}
\usepackage{balance}
\usepackage{subcaption}
\usepackage{epstopdf}
\AppendGraphicsExtensions{.pdf}
\usepackage{array}
\usepackage{listings}
\definecolor{Gray}{gray}{0.9}
\definecolor{pgreen}{rgb}{0,0.5,0}
\usepackage{amsthm}
\makeatletter
\def\th@plain{\thm@notefont{}\itshape }
\def\th@definition{\thm@notefont{}\normalfont }
\makeatother
\usepackage{capt-of}

\usepackage{xcolor}
\definecolor{todocolor}{rgb}{0.9,0.1,0.1}
\definecolor{indiagreen}{rgb}{0.07, 0.53, 0.03}
\definecolor{hycolor}{rgb}{0.7,0.7,0.3}
\definecolor{darkbrown}{rgb}{0.4, 0.26, 0.13}

\lstnewenvironment{bash}[1][]
  {\lstset{language=[latex]TeX}\lstset{numbers=left,numberstyle=\scriptsize,
commentstyle=\color{pgreen},
   xleftmargin=5pt,
   xrightmargin=5pt,
   aboveskip=\bigskipamount,
   belowskip=\bigskipamount, #1,
}}
{}

\lstdefinestyle{testlstcolor}{
    language={sh},
    moredelim=**[is][\color{red}]{~}{~},
    moredelim=**[is][\color{blue}]{<}{>},
    moredelim=**[is][\bfseries]{***}{***},
    moredelim=**[is][\color{green}]{~~}{~~},
    showstringspaces=false,
    basicstyle=\ttfamily, 
    literate={\\~}{{\textasciitilde}}1
        {\\<}{{\unichar{"003C}}}1 
        {\\>}{{\unichar{"003E}}}1
}

\newcommand{\tool}{Etor}
\newcommand{\unethicalissues}{1235}\newcommand{\unethicalissuesfinal}{316}\newcommand{\unethicalissuesfinalrepo}{301}\newcommand{\unethicalissuesrepo}{842}\newcommand{\disagreecount}{39}
\newcommand{\disagreepercentage}{12}
\newcommand{\disagreeartifactcount}{28}
\newcommand{\disagreeperartifactcentage}{8.9}
\newcommand{\unethical}{unethical behavior}
\newcommand{\cunethical}{Unethical Behavior}

\newcommand{\categories}{15}
\newcommand{\artifactcategories}{18}\newcommand{\principles}{six}
\newcommand{\noattribute}{No attribution to the author in code}
\newcommand{\lsoftforking}{soft forking}
\newcommand{\softforking}{Soft forking}
\newcommand{\plagiarism}{Plagiarism}
\newcommand{\nolicense}{No license provided in public repository}
\newcommand{\licesIncompatibility}{License incompatibility}
\newcommand{\uninformedlicensechange}{Uninformed license change}
\newcommand{\dependingonproprietarysoftware}{Depending on proprietary software}
\newcommand{\selfpromotion}{Self-promotion}
\newcommand{\lselfpromotion}{self-promotion}
\newcommand{\unmaintaineprojectwithpaidservice}{Unmaintained project with paid service}
\newcommand{\vulnerablecodeAPI}{Vulnerable code/API}
\newcommand{\namingconfusion}{Naming confusion}
\newcommand{\closingissueprwtihoutexplanation}{Closing issue/PR without explanation}
\newcommand{\offensivelanguage}{Offensive language}
\newcommand{\nooptinoptionallowed}{No opt-in or no option allowed}
\newcommand{\privacyviolation}{Privacy Violation}
\newcommand{\propertiescountinGHRepo}{11}

\newcommand{\autodetect}{six}
\newcommand{\guidelines}{11}
\newcommand{\falsepositive}{24.8}
\newcommand{\truepositive}{74.8}

\newcommand{\unmaintainmonth}{0.5}

\newcommand{\evaluationissues}{195,621}
\newcommand{\evaluationrepo}{1,765}
\newcommand{\totalevaluationcorrectdetection}{548}\newcommand{\feedbackreplies}{83}
\newcommand{\feedbackrate}{21.17}
\newcommand{\fixedbeforereport}{19}
\newcommand{\reportedcases}{392}\newcommand{\negativefeedback}{15}
\newcommand{\negativefeedbackrate}{18.07}
\newcommand{\deletedfeedback}{7}
\newcommand{\botfeedback}{5}
\newcommand{\notopenfeedback}{1}

\newcommand{\fixedfeedback}{39}
\newcommand{\fixedfeedbackrate}{57.35}
\newcommand{\positivefeedback}{68}
\newcommand{\positivefeedbackrate}{81.93}
\newcommand{\acceptedfeedback}{29}
\newcommand{\acceptedfeedbackrate}{42.65}

\newcolumntype{L}[1]{>{\raggedright\let\newline\\\arraybackslash\hspace{0pt}}m{#1}}
\newcolumntype{R}[1]{>{\raggedleft\let\newline\\\arraybackslash\hspace{0pt}}m{#1}}

\usepackage{amsmath}
\usepackage{wasysym}
\usepackage{booktabs}

\providecommand\rightarrowrhd{\relbar\joinrel\mathrel\rhd}
\providecommand\longrightarrowRHD{\relbar\joinrel\relbar\joinrel\mathrel\RHD}
\providecommand\longrightarrowrhd{\relbar\joinrel\relbar\joinrel\mathrel\rhd}

\makeatletter
\providecommand*\xrightarrowRHD[2][]{\ext@arrow 0055{\arrowfill@\relbar\relbar\longrightarrowRHD}{#1}{#2}}
\providecommand*\xrightarrowrhd[2][]{\ext@arrow 0055{\arrowfill@\relbar\relbar\longrightarrowrhd}{#1}{#2}}
\makeatother

\AtBeginDocument{\providecommand\BibTeX{{\normalfont B\kern-0.5em{\scshape i\kern-0.25em b}\kern-0.8em\TeX}}}

\begin{document}

\title{Automatic Detecting \cunethical{} in Open-source Software Projects}

\begin{abstract}
Given the rapid growth of Open-Source Software (OSS) projects, ethical considerations are becoming more important. Past studies focused on specific ethical issues (e.g., gender bias and fairness in OSS). There is little to no study on the different types of \unethical{} in OSS projects. We present the first study  of \unethical{} in OSS projects from the stakeholders' perspective. Our study of \unethicalissuesfinal{} GitHub issues provides a taxonomy of \categories{} types of \unethical{} guided by \principles{} ethical principles (e.g., autonomy). Examples of new \unethical{} include \emph{\lsoftforking} (copying a repository without forking) and \emph{\lselfpromotion{}} (promoting a repository without self-identifying as contributor to the repository). We also identify \artifactcategories{} types of software artifacts affected by the \unethical{}. The diverse types of \unethical{} identified in our study (1) call for attentions of developers and researchers when making contributions in GitHub, and (2) point to future research on automated detection of \unethical{} in OSS projects. Based on our study, we propose \tool{}, an approach that can automatically detect \autodetect{} types of \unethical{} by using ontological engineering and Semantic Web Rule Language (SWRL) rules to model GitHub attributes and software artifacts. Our evaluation on \evaluationissues{} GitHub issues (\evaluationrepo{} GitHub repositories) shows that \tool{} can automatically detect \totalevaluationcorrectdetection{} \unethical{} with \truepositive{}\% average true positive rate. This shows the feasibility of automated detection of \unethical{} in OSS projects. \end{abstract}

\author{Hsu Myat Win}
\affiliation{%
  \institution{Southern University of Science and Technology\country{China}}
 }
\email{11960003@mail.sustech.edu.cn}

\author{Haibo Wang}
\affiliation{%
  \institution{Southern University of Science and Technology\country{China}}
  }
\email{wanghb2020@mail.sustech.edu.cn}

\author{Shin Hwei Tan}
\authornote{corresponding author}
\affiliation{\institution{Southern University of Science and Technology\country{China}}}
\email{tansh3@sustech.edu.cn}

\maketitle

\section{Introduction}
\label{sec:introduction}
With the increasing popularity of Open-Source Software (OSS) development, ethical considerations have become an important yet often neglected topic within the research community. For example, the incident where researchers investigated the feasibility of stealthily introducing vulnerabilities in OSS by making \emph{hypocrite commits} (commits that deliberately introduces critical bugs into code), has provoked active discussion among the Linux community, researchers, and other OSS developers~\cite{wu2021feasibility}. The Linux developers argued that making ``hypocrite commits'' is ``not ethical'', and wasting developers' time in reviewing invalid patches~\cite{linuxhypo}. More importantly, this incident has revealed an attack on the basic premise of OSS itself (i.e., the fact that anyone can contribute to the code and any OSS project is susceptible to a similar incident). Indeed, \unethical{} committed by OSS contributors might lead to broken trust between the OSS community and the contributor, whereas unethical software development might lead to loss of funding, reputation, or other resources of the OSS organization involved. Despite the importance of understanding the unethical issues  by \emph{stakeholders} (individuals who participated or interested in the OSS project, and can either affect or be affected by the OSS project), most studies on \unethical{} in OSS projects mainly focuses on the common types of \unethical{}, such as gender bias~\cite{terrell2016gender,InvestigatingBiasGitHub}, fairness in the code review process~\cite{german2018my}, and software licensing~\cite{lerner2005scope,vendome2017machine,LicenseUsageChangesJava}. \emph{There is little to no study that investigates the important question: ``What kind of behavior is considered unethical by stakeholders in OSS projects?''}. Without understanding the definition of \unethical{} from the perspective of the stakeholders of OSS projects, incidents similar to the ``hypocrite commits'' experiments are bound to reoccur. 

Prior studies stress the importance of considering ethical issues in OSS projects by using various examples and referring to ethical principles~\cite{ResearchEthicsOpenSource,EthicalIssuesInOSS,EthicalMining}. Unfortunately, a study revealed that instructing participants to consider the ACM code of ethics does not affect their ethical decision-making in software engineering tasks~\cite{ACMChanges}. A similar argument has been made in AI ethics, which calls for practical methods to translate principles into practice~\cite{aiaethical}. Hence, we argue that it is \emph{not enough to merely observe the occurrence of \unethical{} via several examples in OSS projects}, it is much more important to \emph{systematically study their characteristics, and design practical tools that can automatically detect \unethical{} by presenting evidences using data in OSS projects to stakeholders.}

To bridge the gaps between general ethical principles and OSS practices, we present the \emph{first study of the types of \unethical{} in OSS projects from stakeholders' perspectives}. Specifically, our study aims to answer the research questions below:

\noindent
\textit{\textbf{(RQ1) How does stakeholder in OSS projects define \unethical{}, and what are the types of \unethical{}?}}

By referring to \emph{ethical principles}, we study the \emph{diverse types of \unethical{}}, their \emph{characteristics}, and the corresponding \emph{ethical principles} that drive these \unethical{} in OSS projects.

\noindent\textit{\textbf{(RQ2) Given the type of \unethical{}, what is the corresponding type of software artifacts that are deemed as unethical by stakeholders of OSS project?}}

For each type of \unethical{}, we study the \emph{affected software artifacts} (i.e., artifacts in which the stakeholders claimed to violate ethical principles) to guide our design of a tool that can automatically identify \unethical{} in OSS projects.

Our study leads to a taxonomy of \categories{} types of \unethical{} in OSS projects. including S1: \noattribute{}, S2: \softforking{}, S3: \plagiarism{},  S4: \licesIncompatibility{}, S5: \nolicense{}, S6: \uninformedlicensechange{}, S7: \dependingonproprietarysoftware{}, S8: \selfpromotion{}, S9: \unmaintaineprojectwithpaidservice{}, S10: \vulnerablecodeAPI{}, S11: \namingconfusion{}, S12: \closingissueprwtihoutexplanation{}, S13: \offensivelanguage{}, S14: \nooptinoptionallowed{}, and S15: \privacyviolation{}. Six of them have not been studied (i.e., S2, S6, S8, S9, S11, S12). For example, our study discovered the \unethical{} of ``S8: \selfpromotion{}'' where a contributor $C$ deliberately opened many new pull requests $PR$s in several popular OSS projects where the code of the $PR$s depends on a newly released library $L$ in which $C$ is a contributor without mentioning the conflict of interests (the fact that he is promoting his own library)~\cite{S8}. Another example is ``S11: \namingconfusion{}'' where the developer selects a conflicting name for an artifact which is the same as existing names but stakeholders should be responsible for selecting unique names.

Inspired by our study, we propose \underline{E}thic detec\underline{tor} (\tool{}), an automatic detection tool based on \emph{ontological engineering} (a description of entities and their properties, relationships, and behaviors) and Semantic Web Rule Language (SWRL) rules to model software artifacts in GitHub. In summary, we made the following contributions: \begin{itemize}[leftmargin=*,labelindent=3pt,nosep]
    \item\textbf{Study.} To the best of our knowledge, we conducted the first study of \unethical{} in OSS projects from the stakeholders' perspective. Our study of \unethicalissuesfinal{} GitHub issues/PRs from \unethicalissuesfinalrepo{} projects revealed \categories{} types of \unethical{} with six new ones.
Our study also revealed the diversity of the affected software artifacts. Our benchmark containing \unethicalissuesfinal{} issues with various types of \unethical{} lays the foundation for future automated approaches for detecting \unethical{}. \item \textbf{Technique.} We propose \tool{}, a novel ontology-based tool that automatically detects \unethical{} in OSS projects. We model GitHub attributes using ontologies, and design SWRL rules to check for \unethical{} in various artifacts. \item \textbf{Evaluation.} Our evaluation on \evaluationissues{} GitHub issues/ PRs from \evaluationrepo{} repos shows that \tool{} can automatically detect \totalevaluationcorrectdetection{} issues with \truepositive{}\% true positive rate on average. \end{itemize}

 \section{Background And Related work}
\label{sec:backgroundAndRelatedWorks}

\noindent\textbf{\emph{Ethical Principles.}} 
Prior work on ethical principles in OSS projects mainly studied \principles{} aspects: (1) accountability, (2) attribution, (3) autonomy, (4) informed consent, (5) privacy, and (6) trust~\cite{InformationTransparency,ValueSensitive,rightThingOnline,Crowdsourcing}. \textbf{Accountability} means that an individual is accountable for his/her actions. \textbf{Attribution} (e.g., copyright) means giving credit to authors when the credit is due. \textbf{Autonomy} allows an individual to decide, plan, and act to achieve their goals. In OSS projects, individuals inherently have
autonomy because they can choose which tasks to perform but may gain or lose autonomy once they agree to participate. \textbf{Informed Consent} is an agreement between the individual and the institution maintaining ethical values, such as autonomy. \textbf{Privacy} is a right of a stakeholder on what information another stakeholder can obtain and communicate to others. \textbf{Trust} refers to expectations between people through goodwill.

\noindent\textbf{\emph{Web Ontology Language (OWL)}} is a standard ontology language endorsed by the W3C to construct an OWL knowledge model~\cite{RN249,mcguinness2004owl,antoniou2004web}. It is a semantic web language designed to model rich and complex knowledge about things, groups of things, and relations between things. Knowledge expressed in OWL can be exploited by computer programs, e.g., to verify the consistency of that knowledge or to make implicit knowledge explicit.
Thus, we design our tool based on ontology engineering.

\noindent\textbf{\emph{Semantic Web Rule Language (SWRL)}} is a language that combines OWL and Rule Markup Language (RuleML), which can be used to express Horn-like rules and logic~\cite{RN250}. SWRL rules are used to infer new knowledge regarding the individual (instance) by chains of properties. We choose to model the \unethical{} in OSS projects using SWRL because (1) its expressiveness~\cite{vrandevcic2009ontology} is well-suited for modeling \unethical{} that involves different GitHub attributes and diverse types of software artifacts, and (2) it has been widely used to model concepts such as privacy for medical data~\cite{boussi2009modelling} and access control policy~\cite{kayes2018accessing,beimel2011using}.

\noindent\textbf{\emph{Related Work.}} Prior work studies focus on multiple aspects of ethical concerns for several domains.

\noindent\emph{Ethical concerns in Software Engineering research.} Several studies focus on ethical concerns for empirical studies in software engineering. Badampudi conducted a study about the reports of the ethical considerations in Software Engineering publications~\cite{ReportingEthics}. Andrews et al. illustrated some of the common approaches to encourage ethical behavior and their limits for demanding ethical behavior between researchers' duty and their publishing as well as the companies' and individuals' integrity~\cite{LimitsOfPolicy}. Singer et al. introduced their work as a practical guide to ethical research involving humans in software engineering~\cite{EthicalIssuesEmpiricalStudiesSE}. Our study is complementary to these studies as the types of \unethical{} discovered in our study points to potential violations of ethical principles that software engineering researchers should consider when their evaluations of automated tools use OSS projects.  

\noindent\emph{Studies on ethical concerns in OSS.} Existing studies of OSS projects focus on issues related to gender bias~\cite{terrell2016gender,InvestigatingBiasGitHub}, fairness of the code review process~\cite{german2018my}, similar code in Stack Overflow and GitHub~\cite{yang2017stack,baltes2019usage}, and software licensing~\cite{lerner2005scope}~\cite{vendome2017machine}~\cite{LicenseUsageChangesJava}.
Studies relating to gender bias in GitHub~\cite{terrell2016gender,InvestigatingBiasGitHub} aims to address the obstacles in improving gender diversity. Meanwhile, a study of a large industrial open source ecosystem (OpenStack) shows that unfairness is ``starting to be perceived as an issue'' in OSS~\cite{german2018my}. Several studies investigated code clones between code snippets from Stack Overflow and projects on GitHub and found a considerable number of non-trivial clones~\cite{yang2017stack,baltes2019usage}. Although these studies also explored how GitHub stakeholder's reference code was copied or adapted from Stack Overflow answers without giving proper credits to the authors (who wrote the code), they did not consider the scenario where the stakeholder of the code snippets used in GitHub is the same as the owner of the code in Stack Overflow (in this case, a credit is not needed). Several techniques have been proposed for the automated detection of license incompatibility~\cite{german2010sentence,kapitsaki2017automating,xu2021lidetector}. While our study identifies license incompatibility as an \unethical{}, it includes more diverse types of issues related to licensing (e.g., missing license, and uninformed license change). Nevertheless, all existing studies on ethical concerns in OSS projects only focus on a few aspects of ethical principles, and they did not conduct analysis of the diverse types of ethical violations in OSS projects in GitHub.

 \section{Study of \unethical{} in OSS}
\label{subsec:typesofunethicalissues}
To address the two research questions introduced in Section~\ref{sec:introduction}, we conducted a study of \unethical{} in OSS projects. Although using a mixed-method research methodology (e.g., adding a survey that asks developers for their opinions on each \unethical{}) would provide stronger empirical evidences, we choose to observe \unethical{} passively by reading developers' discussions to avoid spamming developers~\cite{baltes2016worse}. \begin{figure}[htbp]\centering
\includegraphics[width=0.9\linewidth]{resource/StudyWithoutFix-pdf-converted-to.png}
\caption{Overview of our study on \unethical{}.}
\label{Figure:emphiricalstudy}
\end{figure}

\noindent{}\textbf{Study methodology.} Figure~\ref{Figure:emphiricalstudy} gives an overview of our study. We built a crawler that crawls GitHub issues by searching using the keyword ``ethic'', concepts related to unethics, and synonyms for ``un/ethical'' (i.e., ``unprofessional'',  ``unfair'', ``right'', ``proper'', and ``principle'') via the GitHub API. We then manually checked the results to exclude issues that do not have a clear description or are unrelated to ethical behavior. After getting the relevant issues, we manually analyzed the stakeholders' discussions using thematic analysis~\cite{cruzes2011recommended}, an approach for identifying patterns (or ``themes'') within data. Specifically, the first two authors of the paper followed five steps: (1) we carefully read and analyzed all discussions in the issue to understand what stakeholders discussed about and how they described \unethical{}, identifying the key sentences and phrases which represent \unethical{}. (2) We coded the key sentences and phrases in each issue by highlighting sections of text, and coming up with shorthand labels or “codes” to describe their content. We reread the related key sentences, phrases, and their surrounding context discussions to generate initial codes. New codes can be added as we go through the discussions. After we have been through the discussions, we collate together all the key sentences and phrases into groups identified by codes. These codes allow us to gain a condensed overview of the main points and common meanings that recur throughout the discussions. (3) After generating initial codes, we looked over the created codes, aggregated codes with similar meaning into groups, and started coming up with themes for those groups. Themes are generally broader than codes. (4) With the initial set of themes in the previous step, we reviewed all themes to look for chances to merge similar themes or sub-theme. (5) We finalized the themes by providing clear definitions. To reduce research bias, steps (1) to (4) were conducted independently by the first two authors of the paper. Then, a sequence of meetings was held to resolve conflicts and define the final themes in step (5). Both authors are PhD students with more than two years of research experience. The first author had taken a computer ethics course, while the second author had experience in OSS development. For RQ1, we develop a taxonomy of the types of \unethical{} in OSS projects and its underlying principles. Before following the steps of thematic analysis, we reviewed ethical principles from prior studies ~\cite{InformationTransparency,ValueSensitive,rightThingOnline,Crowdsourcing}, and identified \principles{} ethical principles guiding the action of stakeholders in OSS projects, including: (1) accountability, (2) attribution, (3) autonomy, (4) informed consent, (5) privacy, and (6) trust (i.e., we exclude ``welfare'' because it is related to fair wages which is generally not discussed in our studied issues). We use these \principles{} underlying ethical principles and their corresponding ethical guidelines as guidance for merging relevant themes. For RQ2, we first obtained the initial ``themes'' (i.e., software artifacts) based on prior work~\cite{pfeiffer2020constitutes,huq2019understanding}. Then, via an iterative process of (1) reading \unethicalissuesfinal{} issues with their corresponding types of \unethical{}, and (2) refining the themes via thematic analysis, we derived \artifactcategories{} types of affected software artifacts.  \begin{figure}[htbp]\centering
\includegraphics[width=0.9\linewidth]{resource/Taxonomy-withoutFix-pdf-converted-to.png}
\caption{Taxonomy of \unethical{} in OSS projects.}
\label{Figure:figuretaxonomy}
\vspace{-0.4cm}
\end{figure}
\begin{table}[h]\centering
\caption{Types and affected artifacts for \unethical{}}\label{tab:taxoCountArtifact}
\footnotesize 
\setlength{\tabcolsep}{2.8pt}
\resizebox{0.98\linewidth}{!}{
\begin{tabular}{|l|R{3em}|p{25em}|} 
\hline
\textbf{Type} & \textbf{Issues (\#)} & \textbf{Affected Software Artifacts}\\ \hline
S1	&	49 & 41 Source code, 5 Configuration files, 1 API, 1 project, 1 script \\ \hline
S2	&	19 & 19 Projects \\ \hline
S3	&	6 & 2 Source code, 2 Data, 1 UI, 1 Project \\ \hline
S4	&	26 & 9 Legalese, 7 Source code, 4 \texttt{README}/ \texttt{CONTRIBUTING.md}, 3 Configuration files, 1 Image, 1 OS, 1 Website \\\hline
S5	&	31 & 31 Legalese \\ \hline
S6	&	9 & 9 \texttt{CHANGELOGs} \\\hline
S7	&	16 & 16 APIs \\\hline
S8	&	8 & 8 PR/Issue comments \\\hline
S9	&	10 & 10 Release histories \\\hline
S10	&	27 & 23 Source code, 4 APIs \\\hline
S11	&	21 & 10 Product names, 8 Source code, 1 UI, 1 Data, 1 Script\\\hline S12	&	15 & 10 PR/Issue comments, 5 PR/Issue code reviews \\\hline
S13	&	7 & 2 UIs, 2 Product names, 1 Source code, 1 \texttt{README}/ \texttt{CONTRIBUTING.md}, 1 Website\\\hline
S14	&	36 & 15 UIs, 11 Software features, 6 Source code, 4 Configuration files \\\hline
S15	&	36 & 12 Source code, 10 APIs, 5 UIs, 5 Software features, 3 Configuration files, 1 Website \\\hline
\hline
\textbf{Total} & \textbf{\unethicalissuesfinal{}} & \textbf{\unethicalissuesfinal{}}\\\hline
\end{tabular}}
\end{table}
 \subsection{RQ1: Types of \unethical{}}
\label{subsection:RQ1}
We crawled issues in GitHub, and obtained \unethicalissues{} issues/PRs of \unethicalissuesrepo{} projects submitted by stakeholders. After reading the stakeholders' discussion in GitHub issue/PRs and manually filtering out the invalid issues (e.g., issues that mentioned ``ethic''' but only involved updating terms and conditions in document~\cite{InvalidIssue1}), we obtained \unethicalissuesfinal{} issues with 23 keywords (e.g., ``copy'', ``plagiarism'') shown in the supplementary material. We then identified themes in these keywords by referring to the \principles{} principles and their corresponding guidelines. For example, keywords such as ``copy'' and ``plagiarism'' belong to the same ethical guideline (``To respect copyright'') for ``(S1) \noattribute{}'', ``(S2) \softforking{}'', and ``(S3) \plagiarism{}'' as they are all related to giving proper credits to the authors but we separate them into different types as they involve different degrees of copying (copying entire repository in ``\softforking{}'' versus copying texts in ``\plagiarism{}''). Subsequently, we obtained \categories{} types of \unethical{} with \guidelines{} ethical guidelines. After the generation of initial themes, both authors meet to discuss the \disagreecount{} cases (\disagreepercentage{}\%) with divergent themes to reach a consensus.

Figure~\ref{Figure:figuretaxonomy} shows the \categories{} types of \unethical{} in our study. Boxes on the left (e.g., ``Attribution'') describes the ethical principle behind each type, whereas the grey heading for the boxes on the right (e.g., ``To respect copyright'') includes the \guidelines{} ethical guidelines, and the contents present the related types of \unethical{}. Six of the \categories{} types have not been previously studied (i.e., S2, S6, S8, S9, S11, S12).
\begin{tcolorbox}[left=1pt,right=1pt,top=1pt,bottom=1pt]
\vspace{-3pt}
\textbf{Finding 1:} The types of \unethical{} in OSS projects are diverse (\categories{} types identified in our study, with six new ones) \vspace{-3pt}
\end{tcolorbox}

We explain the \guidelines{} ethical guidelines and the corresponding types of \unethical{} below:

\begin{enumerate}[wide,labelindent=0pt,nosep]	
\item \emph{To respect copyright.} There are three types of \unethical{} related to copyright, described below:

\noindent \textbf{S1: \noattribute{}.} This issue occurs when the stakeholders failed to give proper credit after copying a piece of code~\cite{baltes2017attribution}. An example for \textbf{S1} is: 

\emph{(S1) ``it is \textbf{unethical} not to credit or at the least, point out that these features are inspired by...''~\cite{S1}}

\noindent \textbf{S2: \softforking{}.} This issue occurs if the copied item is a \emph{repository} and the copied repository has not been forked. Although GitHub encourages forking for social coding, a copied repository should acknowledge the original repository by creating an official fork~\cite{nyman2011fork}.  An example discussion for \textbf{S2} is: 

\emph{(S2) ``Unauthorised copy of... \textbf{unethical}... You must delete this repo and fork from the original...''~\cite{S2}}

\noindent \textbf{S3: \plagiarism{}.}
\plagiarism{} occurs if the stakeholders copied texts (non-source code) or the entire product regardless of giving credit or not~\cite{RN295}. An example discussion for \textbf{S3} is:

\emph{(S3) ``Interactive book should be free of plagiarism. By replicating the content used by...\textbf{unethical}.''~\cite{S3}}  

In this example, the repository of an interactive book is unethical because the book uses copied texts from several websites. 

\item \emph{To help individuals make informed consent decisions easier via licensing.} There are three types of \unethical{} related to licensing, described below:

\noindent \textbf{S4: \licesIncompatibility{}.}
It occurs if the repository includes source code or text files carrying different license types than the project's license because stakeholders must ensure license compatibility of the repository. Example for \textbf{S4} is: 

\emph{(S4) ``To continue distributing when we know they have incompatible licenses is \textbf{unethical}.''~\cite{S4}}.

\noindent \textbf{S5: \nolicense{}.}
This issue occurs if the public repository does not have any license and the stakeholders request for it because licenses state the official permissions to use a repository, and project owners should provide them if the OSS is public for greater transparency. An example comment for \textbf{S5} is:

\emph{(S5) ``The repository is public which implies an intent of being open-source but no license is specified making review of the code an issue...People get...at the end of the day, but they are funding this stuff instead of the... developers. That's \textbf{unethical} but legal.''~\cite{S5}}. 

\noindent \textbf{S6: \uninformedlicensechange{}.} 
Due to transparency concerns, OSS developers should inform the stakeholders about the license change (via \texttt{CHANGELOG} or PR) prior to changing the license. \textbf{S6} occurs if the contributors fail to do so. An example for \textbf{S6} is:

\emph{(S6) ``Normally license change are announced in some form of PR or announcement or discussion and none of that has taken place...I find this silent change \textbf{unethical}.''~\cite{S6}}

\item \emph{To avoid license violation}. Stakeholders must obey the OSS license agreement and avoid integrating prohibited licenses that cause violations in license dependency chains~\cite{kapitsaki2015insight}. 

\noindent \textbf{S7: \dependingonproprietarysoftware{}.} This issue occurs if the OSS project relies on closed-source software because OSS projects should be fully open-sourced. An example comment for \textbf{S7} is: 

\emph{(S7) ``Since ... is fully open source software, I believe depending on closed source software is \textbf{unethical}''~\cite{S7}}.

\item \emph{To respect expectations
between people through goodwill}. Trust is an ethical principle that refers to respecting expectations between people through goodwill. The following type of \unethical{} may lead to broken trust among stakeholders in OSS projects:

\noindent \textbf{S8: \selfpromotion{}.}
This issue occurs when the stakeholder advertises his or her repository by suggesting to incorporate it into another repository \emph{without mentioning that he or she is a contributor or owner of the artifact}. This goal of the stakeholder is to attract attention to his or her less well-known repository to increase its popularity. An example comment for \textbf{S8} is: 

\emph{(S8) ``I strongly advise against migrating to nanocolors...Seeing him leverage his notability and following to promote and increase the adoption of nanocolors ..., which he just released a few days ago, is \textbf{unethical}...failing to disclose that you are promoting your own package here is a bad''~\cite{S8}}

In this example, an external (not affiliated with the ESLint repository) user who is the owner of the \texttt{nanocolors} library opens a PR in the ESlint repository to suggest replacing the \texttt{chalk} library with \texttt{nanocolors} to promote his own library. To mitigate this issue, a contributor of ESLint later suggested the user to disclose the fact that he is promoting his own library.

\item \emph{To be responsible for the project maintenance.} Project owners, especially those who offer paid services should actively maintain their projects. If the project owners would to discontinue their technical supports, they should inform the users before asking them to pay for the service. 

\noindent \textbf{S9: \unmaintaineprojectwithpaidservice{}.} This issue occurs if the project repository is not actively maintained when it has a paid service. It is unethical because the project owner is responsible for providing support to paid users who reported the bugs, and fix the bugs within a reasonable time. An example for \textbf{S9} is: 

\emph{(S9) ``I just bought the pro version, and now I’m having this same problem...definitely \textbf{unethical}.''~\cite{S9}}

In this example, the user who has paid for the open-source app reported the failure in using themes (a functionality that is only available for paid users) but the app is no longer maintained. 

\item \emph{To avoid fraudulent activities.} As a code of conduct in OSS, stakeholders should be aware of malicious activities.

\noindent \textbf{S10: \vulnerablecodeAPI{}.}
The issue occurs when stakeholders or a project is involved in malicious activities (e.g., contributing malicious code/API or leaving an unfixed vulnerability in the code).
An example comment for \textbf{S10} is:

\emph{(S10) ``Given that iText 2.1.7 has...unfixed security vulnerability, ...continuing to release it is \textbf{unethical}. In my opinion, iText 2.1.7 should be replaced by OpenPDF.''~\cite{S10}}.

In this example, the user suggested replacing \texttt{iText} which has unfixed vulnerability with another library (\texttt{OpenPDF}) where the vulnerability has been fixed. Another example for \textbf{S10} is the ``hypocrite commits'' incident mentioned in Section~\ref{sec:introduction}.

\item \emph{To be responsible for naming.}
Stakeholders are responsible for all software artifacts that they owned, including the selected names. 

\noindent \textbf{S11: \namingconfusion{}.} This issue occurs when it involves the stakeholders' duty to give unique names for their artifacts (e.g., packages, variables, and libraries). Project owners should identify unique names before using the names. An example for \textbf{S11} is:

\emph{(S11) ``There is already a package `click' for creating command-line interfaces. I am using \texttt{coreapi} package which ... import \texttt{click} package:... your library does not have a style component and python throws an error...this kind of behavior for a company... \textbf{unethical}''~\cite{S11}}.

In the above example, a user complained that the developers of the \texttt{click-integration-django} library select the same package name as the \texttt{Click} package, causing a error when using the package due to naming conflicts.

\item \emph{To be responsible for explaining public
actions.} Owners of OSS projects should explain each decision made for supporting users. 

\noindent \textbf{S12: \closingissueprwtihoutexplanation{}.} This occurs when an issue/PR has been closed without providing any explanation because all stakeholders are expected to receive reasonable explanations for  informational fairness~\cite{colquitt2001dimensionality}. An example for \textbf{S12} is:

\emph{(S12) ``It's a bit \textbf{unfair} to just close something without explaining why?...I don't understand why this (despite several closed issues all saying the same thing) isn't being implemented''~\cite{S12}}.

\item\emph{To avoid offensive language.} 
Stakeholders should encourage respectful environment in OSS projects by avoiding offensive language because words with offensive language might represent \unethical{}~\cite{da2021could}. Prior study stated that hate speech (offensive words) might not be a criminal offense but can still be harmful~\cite{mondal2017measurement}.

\noindent \textbf{S13: \offensivelanguage{}.} This occurs if the stakeholders or part of the project uses offensive language. An example for \textbf{S13} is:

\emph{(S13) ``Rename the Scroll of Genocide to something else...It was never a good or \textbf{ethical} name...It is not ``merely'' systemic and deliberate mass-murder...but state-enacted systemic destruction, neglect and suppression of entire schools of culture, science, literature, truth, of everything that makes us human''~\cite{S13}}. 

In this example, the stakeholder thinks that using the word ``Genocide'' to name a scroll in the open-source game is unethical because the word promotes intentional destruction of human being. \item \emph{To allow individuals to choose which tasks to perform.} Based on the ``Autonomy'' ethical principle, stakeholders of OSS should have the freedom to choose the tasks to perform.

\noindent \textbf{S14: \nooptinoptionallowed{}.} 
This occurs if the system does not provide users options such as withdrawing from using the product. For example, no option is available for uninstalling the third-party library. We focus on issues with ``no option'' or ``no opt-in'' because they provide stronger protections than opt-out~\cite{bergerson2000commerce}. 
An example comment for \textbf{S14} is:

\emph{(S14) ``There should be an option if someone wants to completely remove ... from the system...I think it's \textbf{unethical} to not provide an easy way for a program to be uninstalled''~\cite{S14}}. \item \emph{To protect the right of an individual of personal information}. The privacy of stakeholders of OSS should be protected.
 
 \noindent \textbf{S15: \privacyviolation{}.} 
This occurs in OSS projects under two common scenarios: (1) if the software still collects data despite opting-out via consent, and (2) if there exist personal data leaks regardless of the options (opt-in/out). Example for \textbf{S15} is:

\emph{(S15) ``Form submitted even if opt-in checkbox is unchecked...Signing people up when they haven't opted in is a major enough bug that it renders the plugin useless (or at least \textbf{unethical})''~\cite{S15}}. \end{enumerate}Table~\ref{tab:taxoCountArtifact} presents the numbers of issues we found for each type of \unethical{}. The ``Type'' and ``Issues (\#)'' columns represent the types of \unethical{} and the number of issues we found in GitHub, respectively. Overall, our study identifies \categories{} types of \unethical{} where \emph{the most common types of \unethical{} are related to copyright (\textbf{S1}, \textbf{S2}, and \textbf{S3}) and licensing (\textbf{S4}, \textbf{S5}, \textbf{S6}, and \textbf{S7})}. As our study shows that illegal copying of code (\textbf{S1}) or copying the entire repository (\textbf{S2}), or copying texts (\textbf{S3}) are common in OSS projects, we hope to raise awareness to stakeholders of OSS projects that such behavior is considered unethical.
\begin{tcolorbox}[left=1pt,right=1pt,top=1pt,bottom=1pt]
\vspace{-3pt}
\textbf{Finding 2:} The most common types of \unethical{} in OSS are issues related to copyright (23\%) and licensing (26\%).
\vspace{-3pt}
\end{tcolorbox}

\subsection{RQ2: Affected software artifacts}
\label{subsec:rq2}
We define \emph{affected software artifacts} as objects in software repositories that violate ethical principles. To derive the set of affected software artifacts, we started with the 19 categories from the taxonomy of prior study~\cite{pfeiffer2020constitutes}. Then, we categorized the artifacts we found in our study based on the 19 categories. After removing categories with no artifact found, we obtained eight categories: (1) source code, (2) script, (3) configuration, (4) database (data), (5) image, (6) prose, (7) legalese, and (8) other. For the \emph{prose} category (i.e., plain text files), we only found two concrete types (i.e., README/CONTRIBUTING.md, and CHANGELOG) so we separated them into two categories. As the category ``other'' in prior study~\cite{pfeiffer2020constitutes} is too broad, we split it into 10 new categories based on aforementioned steps in thematic analysis: (1) external application programming interface (API), (2) user interface (UI), (3) project, (4) release history, (5) software feature, (6) product name, (7) operating system (OS), (8) website, (9) PR/Issue code review, (10) PR/Issue comment. We derive ``PR/Issue code review'' and ``PR/Issue comment'' based on prior work~\cite{huq2019understanding}. Our newly introduced categories aim to preserve the hierarchy of artifacts (Project $>$ Software feature~\cite{eisenbarth2003locating} $>$ Source code). For \disagreeartifactcount{} cases (\disagreeperartifactcentage{}\%), both authors meet to discuss the issues labeled with different categories to resolve any disagreement.  Finally, we obtained \artifactcategories{} types of affected software artifacts: (1) project, (2) software feature, (3) source code, (4) external API, (5) legalese, (6) product name, (7) release history, (8) UI, (9) configuration file, (10) PR/Issue code review, (11) PR/Issue comment, (12) README / CONTRIBUTING.md, (13) CHANGELOG, (14) data, (15) image, (16) OS, (17) website, and (18) script (i.e., source code in languages executed by an interpreter). As several artifacts are more difficult to understand, we explain them below: 
\begin{description}[wide,labelindent=0pt,nosep]
\item[Project:] The affected artifacts involve more than one types of artifacts within the entire repository.
\item[Software feature:] Functional or non-functional requirements of a system~\cite{eisenbarth2003locating,hsi2000studying}. An example is the ability to unsubscribe a service.
\item[Source code:] Source files (excluding scripts, binary code, build code) that belong to the current repository (internal).
\item[External API:] API from third party (external) library or service.
\item[Legalese:] Licenses, copyright notes, or patents. \item[Product name:]  The product, project, or app name.

\begin{tcolorbox}[left=1pt,right=1pt,top=1pt,bottom=1pt]
\vspace{-3pt}
\textbf{Finding 3:} The \unethical{} in OSS projects affect many different types of software artifacts (our study found \artifactcategories{} types). 
\vspace{-3pt}
\end{tcolorbox}

\end{description}

The third column in Table~\ref{tab:taxoCountArtifact} presents the affected artifacts for each \unethical{}. Each number in the column denotes the number of GitHub issues with a certain type of artifact (e.g., ``19 Projects'' means that there are 19 issues where \textbf{S2} is affected by projects). Theoretically, one issue might discuss multiple artifacts but we found that each issue only discusses one artifact because (1) developers prefer discussing ethical concerns for one type of artifact in one issue, and (2) some categories are hierarchical (e.g., ``project'' includes multiple types of artifacts). Overall, Table~\ref{tab:taxoCountArtifact} shows that \emph{source code is still the most common type of artifacts for \unethical{}} (i.e., it affects eight types of \unethical{}). \begin{tcolorbox}[left=1pt,right=1pt,top=1pt,bottom=1pt]
\vspace{-3pt}
\textbf{Finding 4:} Source code is the most common types of affected artifacts (affects eight types of \unethical{}). \vspace{-3pt}
\end{tcolorbox}

 \section{Methodology}
\label{sec:methodology}

Our study shows that diverse types of \unethical{} exist in OSS projects, and they usually involve diverse types of software artifacts. The diversity and the complexity of the rules governing the ethics-related activities in GitHub motivate the need for a modeling approach that can abstract this complexity and facilitate its automatic detection. In Section~\ref{sec:model}, we describe how we model \unethical{} using SWRL rules. Then, we explain the architecture of \tool{} that uses SWRL rules for automatic detection in Section~\ref{sec:tooldetect}. 

\begin{table}[h]\caption{GitHub attributes and types for auto-detection}
\label{tab:features}
\resizebox{0.98\linewidth}{!}{\begin{tabular}{|l|l|l|}
\hline
\textbf{Attribute} & \textbf{Type} &\textbf{Description} \\
\hline

\multicolumn{3}{|c|}{\texttt{GHRepository:}  main class}\\\hline
licenseFile & GHContent & repo's license file\\\hline
readmeFile & GHContent & readme file\\\hline
fileCount & int & \# of files in repo\\\hline
fileContent & GHContent & file's content\\\hline
commitCountByPath & int & \# of commits for specific file path\\\hline
commitByPath & GHCommit & commit for file path\\\hline fork & GHRepository & fork of a repo\\\hline
forkCount & int & \# of forks of repo\\\hline
contributor & GHUser & stakeholder taking part in GitHub repo\\\hline
pullRequestCountByCommit & int & \# of PRs which contain specific commit\\\hline
latestRelease & GHRelease & the last release in GitHub history\\\hline
\multicolumn{3}{|c|}{\texttt{GHUser:} A GitHub user identified by username}\\\hline
user & String & GitHub username\\\hline
\multicolumn{3}{|c|}{\texttt{GHIssue:} A GitHub issue that describes a bug or a feature.}\\\hline
issueMessageBody & String & description of issue\\\hline
issueOwner & GHUser & stakeholder who reports an issue\\\hline
\multicolumn{3}{|c|}{\texttt{GHCommit:} The code changes in a commit.}\\\hline
commitCodeChange & String & code change in commit\\\hline
\multicolumn{3}{|c|}{\texttt{GHContent:} The content (including source code) of a file and its
location (file path).}\\\hline
contentCount &	int & \# of contents stored in file's content\\\hline
content &	String & content\\\hline
path &	String & file path \\\hline
pathCount &	int & \# of file paths \\\hline

\multicolumn{3}{|c|}{\texttt{GHRelease:} A latest release is represented by the published date of the release}\\\hline
publishedDate & Date & date of release in GitHub history\\\hline
\end{tabular}}
\end{table}

\subsection{Modeling via SWRL rules}
\label{sec:model}
We propose using SWRL rules to represent \unethical{} in an OSS project together with the publicly available data in GitHub. SWRL rules allow us to model affected software artifacts as hierarchies of classes and properties, capturing the relationships between affected software artifacts and stakeholders. Table~\ref{tab:features} shows GitHub attributes used in our modeling. The columns under ``Attribute'', and ``Type'' explain each attribute and its type. 
We model each OSS project as \texttt{GHRepository}. By referring to the GitHub Repositories API~\cite{Repositories}, we selected \propertiescountinGHRepo{} data properties (e.g., \texttt{latestRelease} and \texttt{licenseFile}) that belong to a \texttt{GHRepository} by excluding properties that are irrelevant for \unethical{} (e.g., \texttt{avatar\_url} that points to the icon for a repository). Apart from \texttt{GHRepository}, we introduce six classes to model data properties of a repository: (1) \texttt{GHUser}, (2) \texttt{GHCommit}, (3) \texttt{GHContent}, (4) \texttt{GHIssue} (5) \texttt{GHPullRequest}), (6) \texttt{GHRelease}. While GitHub users (\texttt{GHUser}) usually play different roles in OSS projects, we only model: (1) contributors (users who are official contributors of a repository) and (2) issue owners (users who report an issue). For modeling \texttt{GHIssue}, we reuse the same convention in GitHub by modeling a PR (\texttt{GHPullRequest}) as a subclass of \texttt{GHIssue} (i.e., GitHub Issue Search API will search for issues and PRs, essentially treating a PR as a type of GitHub issue). Figure~\ref{Figure:classeshierarchy} shows the OWL ontology for our model where \texttt{GHRepository} is the main class, and the arrows denote the relationships between the classes. Specifically, $\displaystyle \texttt{GHIssue} \rightarrowrhd \texttt{GHPullRequest}$ represents the subclass relations, whereas other arrows denote \texttt{hasA} relations (e.g., $\displaystyle \texttt{GHIssue} \rightarrowrhd        \texttt{GHUser}$ means that each issue has a user who reports the issue).

\subsection{Automatic detection of \unethical{}}
\label{sec:tooldetect}
We designed \tool{} to auto-detect \autodetect{} types. We excluded nine types because (1) they involve artifacts (e.g., product names, software features) that are difficult to automatically isolate from other artifacts (i.e., ``\nooptinoptionallowed{}'', ``\privacyviolation{}'', ``\namingconfusion{}'', and ``\offensivelanguage{}''), (2) they require sophisticated analysis of configuration files, API or source code  (i.e., ``\plagiarism{}'', ``\dependingonproprietarysoftware{}'', and ``\vulnerablecodeAPI{}''), (3) their detection requires advanced natural language processing (i.e., ``\closingissueprwtihoutexplanation{}'' as it requires automatically checking if the explanation for closing the PR/issue exists), and (4) approaches for ``\licesIncompatibility''~\cite{german2010sentence,kapitsaki2017automating,xu2021lidetector} exist so we exclude it to avoid reinventing the wheels. 
\begin{figure}[htbp]
\begin{center}
\includegraphics[width=0.75\linewidth]{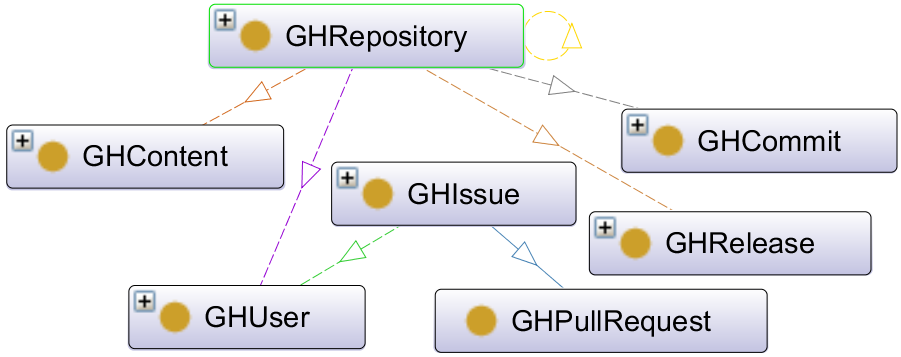}
\caption{Our ontology of \unethical{} in OSS projects}
\label{Figure:classeshierarchy}
\includegraphics[width=0.80\linewidth]{resource/EtorArchi2-pdf-converted-to.png}
\caption{Overall architecture of \tool{} (GH denotes GitHub).} 
\label{Figure:overviewarchitectureofetor}
\end{center}
\end{figure}

\noindent\emph{\textbf{Overview of \tool{}}.} Figure~\ref{Figure:overviewarchitectureofetor} presents the overall architecture of our automatic detection tool, \tool{}. \tool{} supports detection of \unethical{} for two levels, including: (1) repository (denoted as \texttt{repo}), and (2) GitHub issue/pull request (we denote an issue as \texttt{issue} and a pull request as \texttt{PR}).  Given a \texttt{repo} or an \texttt{issue}/\texttt{PR}, and the type of \unethical{} \texttt{eType} to be checked, the \tool{} relies on its set of SWRL rules for its detection, and produces as output whether there is a violation of \texttt{eType} in the given input. Apart from GitHub attributes in Table~\ref{tab:features} that can be detected using the GitHub API, our SWRL rule reasoner uses two additional components for its detection: (1) license detector that checks for licenses at the repository level, and (2) code similarity checker that identifies similar code. 

\noindent\emph{\textbf{Supported  types}.} 
\tool{} supports \autodetect{} types of \unethical{}. 
We include the SWRL rules for all supported types in the supplementary material. We next describe how \tool{} checks each supported type.

\noindent \emph{(S1) \noattribute}. \tool{} checks if an issue or a PR has a Stack Overflow link representing a reference code, and the code snippet copied from Stack Overflow cites the reference link. Although there can be many resources from which stakeholders copy the reference code, \tool{} only check for Stack Overflow links because (1) we learned from our study and from existing work~\cite{baltes2017attribution} that contributors are required to give credit to copied code snippets in Stack Overflows as they are protected by the CC-BY-SA Creative Commons license, and (2) to support other online resources (e.g., GitHub links), we need to automatically extract the original reference code (requires parsing Web pages of different formats), and identify the appropriate license for the code snippet (requires detecting the license for partial code, which is beyond the scope of this paper). Given an \texttt{issue}/\texttt{PR}, \tool{} checks if a comment \texttt{b} in the \texttt{issue}/\texttt{PR} posted by a stakeholder \texttt{u1} contains the Stack Overflow link (\texttt{w}) (we use regular expression to extract \texttt{w}). \tool{} reports a potential violation if: (1) \texttt{u1} is not the owner of the Stack Overflow comment, (2) the code snippets from Stack Overflow is found in one of the files in the repository (\texttt{F}) with at least 10\% similarity (copyright law permits the use of up to 10\% of work without permission~\cite{CopyrightAct1968}), and (3) \texttt{w} is not found in \texttt{F}. 

\noindent \emph{(S2) \softforking{}}. Given two repositories \texttt{r1} and \texttt{r2}, \tool{} compares the contents of all source files in the two repositories to check if one repository is a \emph{soft-fork} (the repository has the same content but it is not listed as an official fork of another repository) of another repository. Specifically, we use \texttt{AC2}~\cite{AC2} to detect the similarities between files. AC2 is a source code plagiarism detection tool that has been widely used by graders to detect plagiarism within a group of assignments. We select AC2 because (1) it supports many programming languages (e.g., C, C++, Java, and PHP), (2) it can be run in a local environment without connection to remote servers, and (3) it is quite robust as it incorporates multiple algorithms found in the literature. \tool{} reports a violation if it detects: (1) 100\% similarity between \texttt{r1} and \texttt{r2}, and (2) \texttt{r2} is not in the fork list of \texttt{r1}. 

\noindent \emph{(S5) \nolicense{}.} Given a repository \texttt{r}, \tool{} detects the repo-level license by checking if it exists in the: (1) \texttt{LICENSE} file~\cite{AddingLicenseToRepository} in the main directory of \texttt{r}, (we check only in the main directory to avoid mistakenly finding API license or package license) or (2) \texttt{README.md} file with license information (we use the list of licenses provided by GitHub~\cite{LicensingRepository} for repo-level license detection). \tool{} reports a potential violation if no license is found after searching for the two files. \begin{table*}[h]
\caption{Number of issues detected and TP/FP rate}
\label{tab:experiment}
\resizebox{0.8\textwidth}{!}{\begin{tabular}{|l|r|rr|rr|r|}
\hline
\multirow{2}{*}{Type}                     & \multicolumn{1}{c|}{\# Unethical Issues} & \multicolumn{2}{c|}{True Positive}                                             & \multicolumn{2}{c|}{False Positive}                                             & \multicolumn{1}{l|}{\multirow{2}{*}{Time (s)}} \\ \cline{2-6}
  & \multicolumn{1}{c|}{\# repos or issues / Total}            & \multicolumn{1}{c|}{\# repos or issues / Total}                        & \multicolumn{1}{c|}{\%} & \multicolumn{1}{c|}{\# repos or issues / Total}                         & \multicolumn{1}{l|}{\%} & \multicolumn{1}{l|}{}                                   \\ \hline
(S1) \noattribute           & 80 / 195,621 issues                          & \multicolumn{1}{r|}{59 / 80 issues}                  & 74                          & \multicolumn{1}{r|}{21 / 80 issues}                   & 26                           & 5.4                                \\ \hline
(S2) \softforking{}                                & 10 / 100 repos                               & \multicolumn{1}{r|}{10 / 10 repos}                   & \textbf{100}                         & \multicolumn{1}{r|}{0 / 10 repos}                       & \textbf{0}                            & 343.1                              \\ \hline

(S5) \nolicense{} & 476 / 1,765 repos                        & \multicolumn{1}{r|}{424 / 476 repos}                 & 89                      & \multicolumn{1}{r|}{52 / 476 repos}                   & 11                      & 3.1                                                     \\ \hline

(S6) \uninformedlicensechange{}                 & 18 / 1,765 repos                         & \multicolumn{1}{r|}{16 / 18 repos}                   & 88                      & \multicolumn{1}{r|}{2 / 18 repos}                     & 11                      & 9.2                                                     \\ \hline

(S8) \selfpromotion{}                                 & 116 / 195,621 issues                         & \multicolumn{1}{r|}{37 / 116 issues}                        & 32                          & \multicolumn{1}{r|}{79 / 116 issues}                         & 68                           & 4.3                                \\ \hline
(S9) Unmaintained Android Project with Paid Service & 3 / 1,765 repos                              & \multicolumn{1}{r|}{2 / 3 repos}                           & \textbf{66}                         & \multicolumn{1}{r|}{1 / 3 repos}                              & \textbf{33}                            & 5.3                                \\ \hline \hline
\textbf{Average}                                & -                         &  \multicolumn{1}{r|}{-}                      & \textbf{\truepositive{}}     &        \multicolumn{1}{r|}{-}             & \textbf{\falsepositive{}}                                                &-                                \\ \hline

\end{tabular}}
\end{table*}

\noindent \emph{(S6) \uninformedlicensechange{}}. We consider a change to be \emph{uninformed} if (1) it is not announced in the  \texttt{CHANGELOG.md} or (2) the license change is not done via PR. Given a repository \texttt{r}, \tool{} checks if the repo-level license has been changed by: (1) extracting commit lists of the license file, and (2) checking if commit changes include license updates. If the license changes occur in more than one commit (we ignore the first commit as it is the initial license creation), \tool{} checks whether the changes have been announced in the \texttt{CHANGELOG.md} by checking whether the \texttt{CHANGELOG.md} mentions license information. If license information is not found, \tool{} checks the PR count for the commit (\texttt{pullRequestCountByCommit}). If the count is less than one, \tool{} marks it as a potential violation. 

\noindent \emph{(S8) \selfpromotion{}}. We consider \emph{self-promotion} to be the scenario where a contributor \texttt{u} opens a GitHub issue/PR where the content of the issue/PR includes links to another repository in GitHub to promote his or her own repository. Given an \texttt{issue}/\texttt{PR} for \texttt{r1} as input, \tool{} first (1) checks that the \texttt{issue}/\texttt{PR} includes a link \texttt{L} to another repository \texttt{r2}, and (2) identifies the stakeholder \texttt{u} who opens the \texttt{issue}/\texttt{PR}. Then, it reports a violation if: (1) \texttt{r1} is not \texttt{r2}, (2) \texttt{u} is not a contributor of \texttt{r1} (i.e., \texttt{u} is an outsider for \texttt{r1}), and (3) \texttt{u} is a contributor of \texttt{r2}. To reduce false positives, \tool{} also checks if \texttt{L} includes specific keywords that usually indicate that the contributor is sharing the link \texttt{L} for demonstration purposes (e.g., \texttt{[DEMO]}) instead of promoting a repository/library (``\textbackslash issues\textbackslash'', ``\textbackslash pull\textbackslash'', ``\textbackslash commit\textbackslash'', ``\textbackslash tree\textbackslash'', ``\textbackslash releases\textbackslash'', ``\textbackslash blob\textbackslash'', and ``\textbackslash runs\textbackslash'').

\noindent \emph{(S9) Unmaintained Android Project with Paid Service}. This type checks whether an Android project offered paid service in Google Play, but stop actively maintaining the GitHub repository. On average, 115 APIs are updated per month~\cite{mcdonnell2013empirical}, and 49\% of app updates have at least one update within 47 days~\cite{mcilroy2016fresh}. Based on this frequency of app updates, we define an \emph{unmaintained Android project} to be an Android project where the latest update is released less than \unmaintainmonth{} year. Given a repository \texttt{r} as input, \tool{} first checks for unmaintained Android projects by examining whether (1) the latest release date (\texttt{D}) of \texttt{r} is less than \unmaintainmonth{} year, and (2) \texttt{r} is an \emph{original repository} (not forked from other repositories). Then, it checks whether the app offers a paid service by (1) identifying the Google Play link \texttt{l} from \texttt{r}, and (2) searching for the ``in-app purchase''. %
 \section{Evaluation}
\label{sec:evaluation}

We applied \tool{} on \evaluationissues{} GitHub issues and PRs of \evaluationrepo{} GitHub repositories to address the following research questions:

\begin{description}[leftmargin=*]
\item[RQ3:] How many unethical issues can \tool{} detect in OSS projects? 
\item[RQ4:] What are the accuracy and efficiency of \tool{} in its detection? \end{description}

By counting the number of unethical issues in OSS projects, RQ3 provides a rough estimation of the prevalence of each type of \unethical{} in OSS projects. 
For RQ4, we measure the accuracy and efficiency of \tool{} using the following metrics: 
\begin{description}[leftmargin=*]
\item[True Positive (TP):] \tool{} classifies an \unethical{} as a potential violation, and it is a true violation.
\item[False Positive (FP):] \tool{} incorrectly classifies an \unethical{} as a potential violation, and it is a false violation.
\item[Time:] The average time taken (in seconds) to detect a type of \unethical{} across all the evaluated repositories/issues.  
\end{description}

\noindent\textbf{Selection of projects/issues.} As there is no prior benchmark for evaluating the detection of \unethical{}, we construct a dataset by crawling GitHub. Our goal is to select the most recent popular (most stars and most forks) OSS projects and the GitHub issues/PRs from OSS projects for evaluation. We first obtain the list of the top 2,000 OSS projects (we first get the top 1,000 projects with the greatest number of stars, and then the top 1,000 projects with the greatest number of forks) created last year (2021). After eliminating duplicated projects, there are \evaluationissues{} GitHub issues/PRs of \evaluationrepo{} projects in our evaluation set. As soft forking requires two repositories as input, we obtain the pair of repositories ($repo1, repo2$) by getting $repo1$ from the top 200 projects (first 100 from most stars, subsequent 100 from most forks) from the initial list of 2,000 projects. From these 200 projects, our crawler automatically identifies $repo2$ by searching GitHub for projects with similar names using the name of $repo1$ as the query. At this step, our crawler found only 10 out of the 200 projects that have repositories with similar names. For each of these 10 projects, our crawler retrieves the first 10 repositories from the search results as $repo2$, leading to a total of 10*10=100 projects for evaluating soft forking.

\noindent\textbf{Ethical considerations.} Before getting feedback from stakeholders, we obtained IRB approval from our institution. As calling out stakeholders for violations of \unethical{} could potentially lead to similar ethical concerns in prior work~\cite{wu2021feasibility}, we choose to evaluate \tool{} by (1) manually inspecting the reported issues, and (2) reporting only the types of \unethical{} with high accuracy (based on our manual analysis). To avoid violating ethical principles as in the ``hypocrite commits'' incident, we explicitly mentioned in each reported issue that we are researchers conducting research on mining software repositories. To reduce author bias in the manual classification of TP/FPs, we ask for help from a non-author to independently label each issue.

All experiments are conducted on a machine with Intel(R) Core (TM) i7-7500 CPU @2.7 GHz and 16 GB RAM.

\noindent\textbf{Implementation.} We use Prot\'eg\'e 5.5.0~\cite{musen2015protege} to define the ontology model. Our crawler uses PyGitHub~\cite{GitHubAPIPython} for querying GitHub.

\subsection{RQ3: Number of detected issues}
\label{subsec:outcomesoftool}
Table~\ref{tab:experiment} summarizes the results of our evaluation. The ``Type'' column denotes the types of \unethical{} detected by \tool{}, whereas the second column is of the form $x$  / $y$ where $x$ represents the number of repositories/issues with the unethical behavior detected and $y$ denotes the total number of repositories/issues in our evaluation dataset. 
Overall, \tool{} has successfully detected at least one violation for all types of \unethical{} that we studied. As our evaluation dataset is different from the study dataset, and we have observed the occurrences of \unethical{} in both datasets, this indicates that \emph{different types of \unethical{} is prevalent in OSS projects}. Table~\ref{tab:experiment} also shows that ``\nolicense{}'' is the most common types among the \autodetect{} types of detected issues. This means that \emph{a relatively high percentage of the evaluated repositories are missing license files} (around 24\% of the evaluated repositories if we exclude the false positives). For the issue-level detection, we observe that ``\noattribute{}'' and ``\selfpromotion{}'' are the most common ones among all evaluated issues/PRs. This indicates that \emph{contributors of OSS projects tend to (1) forget to give credit to the author in their copied code snippets, or (2) promote their own repositories without mentioning they are contributors to the repositories}.

\subsection{RQ4: Accuracy and Efficiency of \tool{}}
\label{subsec:accuracy}
\noindent\textbf{Accuracy.} To evaluate the effectiveness of \tool{}, two raters (one author, and one non-author who is an undergraduate CS student working as a part-time student assistant) independently inspect its output. Specifically, for each violation reported by \tool{}, each rater determines if the violation is a true positive (TP) or a false positive (FP). The initial Cohen’s Kappa was 0.82, which indicates a high level of agreement. The two raters then meet to resolve any disagreement to reach Cohen's Kappa of 1.0. The ``True positive'' and ``False positive'' columns in Table~\ref{tab:experiment} show the results for the inspection. On average, the TP rate is \truepositive{}\%, and the FP rate is \falsepositive{}\%. For repository-level detection, although \tool{} can only detect a small number of violations for ``Soft forking'', it can detect these unethical issues with high accuracy (0\% FP rate). As we consider a repository a \emph{soft-fork} only if all the contents of the two repositories are the same (100\% similarity), this design decision may lead to fewer violations being found but increase the accuracy of its detection. In future, it is worthwhile to study the effect of the similarity threshold on the accuracy of its detection. For issue-level detection, \tool{} can detect S1 with reasonable accuracy (26\% FP rate).

\noindent\textbf{Efficiency.} The ``Time'' column in Table~\ref{tab:experiment} shows the average time taken to detect an \unethical{}. Overall, the average time to analyze a repository is 3.1--343.1 seconds and the average time taken to analyze an issue is 4.3--5.4 seconds. This indicates that \emph{\tool{} can detect a type of \unethical{} relatively fast}. We also observe that ``Soft forking'' is the most time-consuming type to detect because \tool{} needs to check for code similarities for all source files within the repository. 

\noindent\textbf{Reasons behind inaccurate detection.} We manually inspect the reasons behind the FPs reported by \tool{}.
\tool{} reports the highest FP rate for ``\selfpromotion{}''. Recall that \tool{} checks that a stakeholder \texttt{St} opens an issue/PR \texttt{I} at repository \texttt{R1}, and includes the other repository (\texttt{R2}) link (\texttt{L}). A true ``\selfpromotion{}'' only occurs if \texttt{St} did not mention about being a contributor of \texttt{R2}. We need to manually verify this condition by reading the comments written in natural language. Hence, FPs may occur if (1) \texttt{St} mentioned that he or she is a contributor of \texttt{R2} (e.g., ``I am working on a project called the ...'' in comment~\cite{SelfPromotionFalsePositive1}) or (2) \texttt{St} wanted to ask for suggestion in using \texttt{R1} for \texttt{R2} (e.g., ``I'd like to try your ... module in a non-mmdetection repo (...)''~\cite{url25}).

There are three main reasons for FPs in ``\noattribute{}'': (1) no actual copying occurs but a link exists (e.g., the Stack Overflow link was mentioned as references~\cite{url26}), (2) \tool{} checks the exact link and fails to detect if the citation uses the short link of Stack Overflow, and (3) \tool{} matches the exact GitHub user name with the Stack Overflow user name, and fails to detect if the user name is different (e.g., GitHub user name is  \texttt{devinrhode2} and Stack Overflow user name is \texttt{Devin Rhode}~\cite{url27}). For ``\nolicense{}'', FPs occur because the repository (1) has a license file that is not in the main directory (e.g., \texttt{LICENSE} file in the inner folder~\cite{url28}), (2) has a disclaimer in \texttt{README.md} (e.g., ``This repository is for personal study and research purposes only. Please DO NOT USE IT FOR COMMERCIAL PURPOSES.''~\cite{url29}), (3) is used for education purposes (we need to manually exclude repositories for the public schools where the license is not required), (4) has no source code or data, and (5) is under an organization license and no separate license is defined for the repository~\cite{url30}. 
For ``\uninformedlicensechange{}'',  FPs occur because the scenario where the repository has restored the old license should not be considered a violation (e.g., the stakeholder changed the license from ``Apache License Version 2.0'' to ``GNU GENERAL PUBLIC LICENSE Version 3'' on Feb 17, and he/she restored back to ``Apache License Version 2.0'' on Feb 18). For ``Unmaintained Android Project with Paid Service,'' we found one FP because the unmaintained project is a library that an app uses instead of the app itself but the app is actively maintained. (a new version is recently released).

\noindent\textbf{Stakeholders' feedback.} Apart from manually labeling the unethical issues, we also obtained qualitative feedback by reporting them to stakeholders of OSS projects. To avoid spamming OSS developers with inaccurate results, we only reported the types of \unethical{} with >=80\% TP rate in our manual analysis (i.e., \textbf{S2}, \textbf{S5}, \textbf{S6}). For each of these reported types, we opened a GitHub issue to developers when both raters labeled it as TP. We excluded 39 issues because the project owners have disabled GitHub issues (this usually indicates that they do not accept contributions or bug reports~\cite{disableissue}). For example, the repository~\cite{noissuetagsample} violates the ``\nolicense{}'' rule but we cannot report this to the project owner as GitHub issues have been disabled. We also found \fixedbeforereport{} issues that were previously reported and fixed the issues before we file a bug report. In total, we have reported \reportedcases{} issues, and received \feedbackreplies{} replies (a response rate of \feedbackrate{}$\%$) from stakeholders. We carefully looked through all those responses and identified \positivefeedback{} (\positivefeedbackrate{}$\%$) replies as valid and \negativefeedback{} (\negativefeedbackrate{}$\%$) responses as invalid. Among these \positivefeedback{} valid replies, \fixedfeedback{} (\fixedfeedbackrate{}$\%$) have been fixed, and \acceptedfeedback{} (\acceptedfeedbackrate{}$\%$) have been accepted by the stakeholders of the OSS. An example valid feedback that we received is ``Thank you very much for the warning. I have already added the license to the repos that didn't have it.''. For the \negativefeedback{} responses that we considered as invalid,  developers (1) directly deleted or closed our submitted issues without any explanations (\deletedfeedback{}/\negativefeedback{}), (2) thought that the issue reporter is a software bot although we have created the issue manually and explicitly mentioned in the issue that we are researchers (\botfeedback{}/\negativefeedback{}), (3) are not interested in getting any GitHub issues (e.g., claiming that the repository is personal) (2/\negativefeedback{}), and (4) explained that ``Software is not open source but everyone or you can use my soft without license. thank you for support my soft'' (\notopenfeedback{}/\negativefeedback{}).   \section{Discussion and Implications}
\label{discussion}
We discuss practical takeaways and suggestions below:

\noindent \textbf{Implications for stakeholders of OSS projects.} By reading issues that stakeholders in OSS considered as ``\unethical{}'', our study revealed that the types of \unethical{} in OSS projects are diverse (Finding 1), suggesting that \emph{stakeholders of OSS projects should be better educated to create awareness of the different types of \unethical{} when contributing to OSS projects} to avoid violating ethical principles. Apart from general types of \unethical{}, our study also pinpoints six new types of \unethical{} in OSS projects (i.e., (S2) \softforking{}, (S6) \uninformedlicensechange{}, (S8) \selfpromotion{}, (S9) \unmaintaineprojectwithpaidservice{}, (S11) \namingconfusion{}, and (S12) \closingissueprwtihoutexplanation{}). Some of them are related to the unique features of GitHub (e.g., ``\softforking{}'' represents ethical concerns when forking, ``\closingissueprwtihoutexplanation{}'' are related to closing GitHub issues/PRs, ``\selfpromotion{}'' occurs due to the need to promote the popularity of one's new repository, whereas ``\unmaintaineprojectwithpaidservice{}'' denotes the responsibility of an OSS project owner to actively maintain the project to support paid users). The identified new types \emph{call for considerations of the unique context of OSS projects when studying \unethical{}}. Meanwhile, although most software development efforts focus on source code maintenance, our study \emph{urges OSS project owners to be responsible for the product names selection} to avoid violating ``\namingconfusion{}''. As issues related to copyright and licensing are the most common ones (Finding 2), \emph{contributors of OSS projects should pay more attentions in giving appropriate credits, and selecting suitable software licenses} when copying software artifacts or using library. Meanwhile, although source code is the still most common affected software artifacts (Finding 4), our study \emph{urges OSS stakeholders to be responsible for various types of software artifacts} (Finding 3) to avoid violating ethical principles when uploading them to GitHub. 

\noindent\textbf{Implications for researchers and tool designers.} As many of the identified types of \unethical{} (Finding 1) are ethical issues that frequently occur in daily life, our study provides empirical evidence that \emph{there exists some overlaps between the types that occur under general setting} (e.g., ``\plagiarism{}'' and ``\offensivelanguage{}'') \emph{and those that are deemed as \unethical{} by stakeholders in OSS projects}. Indeed, the prevalence of plagiarism is inline with prior study which reported the prevalence of the code borrowing practices in GitHub~\cite{golubev2020study}. Due to the diverse types of \unethical{}, future empirical research should advance beyond the general types of \unethical{}. 

While existing work mostly focus on license incompatibility~\cite{german2010sentence,kapitsaki2017automating,xu2021lidetector}, our study found new issues related to licensing. e.g., ``\uninformedlicensechange{}''. As these issues still occur frequently (Finding 2) and our study identified new types of issues, our study \emph{provides empirical motivations for improving techniques related to copyright and software licenses}. 
 For the newly identified types of \unethical{}, we foresee a huge potential for future research direction in: (1) conducting more \emph{in-depth study in the motivations and the common solutions behind each type of \unethical{}}, and (2) introducing automated techniques that can detect and possibly resolve these issues. We believe that \emph{our taxonomy} of \unethicalissuesfinal{} GitHub issues and \emph{our tool} that uses software artifacts and data available in GitHub API \emph{lay the foundation for future approaches on automated detection} of \unethical{}.  Although source code is still the most common type of affected software artifacts (Finding 4), other artifacts in natural language (e.g., PR/Issue comments, product names, and website) are also common in our study (Finding 3). A promising research direction is to \emph{apply natural language processing techniques to accurately detect affected software artifacts in natural language}. For example, techniques can be designed to automatically \emph{extract and recommend descriptive yet non-conflicting names} (e.g., package names) to avoid ``\namingconfusion{}''. Another future direction is to \emph{design techniques that can automatically identify disclaimer-like statements} to accurately detect ``\closingissueprwtihoutexplanation{}'' (to detect the explanation for the PR/issue), and ``\selfpromotion{}'' (to extract statement where the stakeholder has mentioned being a contributor).

\noindent\textbf{Challenges in automated detection of \unethical{}.} To provide guidelines for future research on the automated detection of \unethical{}, we discuss several challenges identified in our study and evaluation:
\begin{itemize}[wide,labelindent=0pt,nosep]
\item As shown in our study in Section~\ref{subsec:rq2}, the \emph{types of artifacts affected by the \unethical{} are too diverse}. An accurate detection technique needs to support analysis of various types of artifacts, including source code, data, and websites. 
\item Within GitHub, we notice that \emph{discussion and announcement in GitHub spread across multiple web pages} (issues, PRs, wikis, discussions, and commit logs). With the rapid growth of different types of web pages in GitHub, it poses additional challenges for automated approaches to exhaustively analyze all relevant web pages.
\item \emph{Some discussions of \unethical{} occur outside of GitHub} (e.g., personal emails, slack channel). For example, for ``\selfpromotion'', we cannot check whether the stakeholder has communicated with the developers in advance through emails. Without complete information about the discussion, the detection is bound to be inaccurate.
\item \emph{The scope for the detection can be too broad for some types of \unethical{}} (e.g., ``\namingconfusion''). Without a predefined scope of detection (package name collision versus app name collision), we cannot accurately detect the behavior.
\item There exist \emph{ambiguities for certain \unethical{}}, which makes it difficult even for human beings to reach consensus (e.g., whether the language used is offensive). In this case, an automated tool can present all relevant information to help stakeholders in making more grounded ethical decisions. \end{itemize}

 \section{Threats to validity}
\label{threat}

\noindent\textbf{External.} Our findings of \unethical{} may not generalize beyond the studied OSS projects and issues/PRs. There could be \unethical{} that are not reported to the issue tracker. Unfortunately, there is no conceivable way to study these unreported issues. As some issues may not have the ethics-related keywords that we used for searching, we could have also missed some \unethical{}. Nevertheless, our selected keywords already help us in discovering many types of \unethical{}. Hence, we believe the issues in our study provide a representative sample of the reported and resolved unethical issues in our studied repositories. While other types of \unethical{} discovered in our study is important,  \tool{} can only detect \autodetect{} of them, and our evaluation is limited to these \autodetect{} types. Nevertheless, our experiments show that \tool{} can detect \unethical{} with relatively high accuracy.

\noindent\textbf{Internal.} Our code and scripts may have bugs that can affect our results. To mitigate this threat, we make our tool and data publicly available for inspection.

 \section{Conclusion}
To better understand \unethical{} in OSS projects, we conduct a study of the types of \unethical{} in OSS projects. By reading and analyzing the discussion of stakeholders in OSS projects, our study of \unethicalissuesfinal{} GitHub issues identifies \categories{} types of \unethical{}. These \unethical{}s are affected by various types of software artifacts. Inspired by our study, we propose \tool{}, an ontology-based approach that can automatically detect \unethical{}. Our evaluation of \tool{} on \evaluationissues{} issues (\evaluationrepo{} repositories) shows that \tool{} can automatically detect \totalevaluationcorrectdetection{} issues with \truepositive{}\% TP rate on average. As the first study that investigates the types of \unethical{} in OSS projects, we hope to raise awareness among OSS stakeholders regarding the importance of understanding ethical issues in OSS projects. While \tool{} shows promising results in automated detection of \unethical{} in OSS projects, we plan to enhance \tool{} in future to detect more types and reduce false positives using machine learning techniques. %
 
\bibliographystyle{ACM-Reference-Format}
\bibliography{sample-base}


\begin{thebibliography}{83}


\ifx \showCODEN    \undefined \def \showCODEN     #1{\unskip}     \fi
\ifx \showDOI      \undefined \def \showDOI       #1{#1}\fi
\ifx \showISBNx    \undefined \def \showISBNx     #1{\unskip}     \fi
\ifx \showISBNxiii \undefined \def \showISBNxiii  #1{\unskip}     \fi
\ifx \showISSN     \undefined \def \showISSN      #1{\unskip}     \fi
\ifx \showLCCN     \undefined \def \showLCCN      #1{\unskip}     \fi
\ifx \shownote     \undefined \def \shownote      #1{#1}          \fi
\ifx \showarticletitle \undefined \def \showarticletitle #1{#1}   \fi
\ifx \showURL      \undefined \def \showURL       {\relax}        \fi
\providecommand\bibfield[2]{#2}
\providecommand\bibinfo[2]{#2}
\providecommand\natexlab[1]{#1}
\providecommand\showeprint[2][]{arXiv:#2}

\bibitem[\protect\citeauthoryear{??}{S8}{[n.d.]}]%
        {S8}
 \bibinfo{year}{[n.d.]}\natexlab{}.
\newblock
\newblock
\urldef\tempurl%
\url{https://github.com/eslint/eslint/pull/15102}
\showURL{%
\tempurl}


\bibitem[\protect\citeauthoryear{??}{RN2}{[n.d.]a}]%
        {RN249}
 \bibinfo{year}{[n.d.]}\natexlab{a}.
\newblock
\newblock
\urldef\tempurl%
\url{https://www.w3.org/2001/sw/\#owl}
\showURL{%
\tempurl}


\bibitem[\protect\citeauthoryear{??}{RN2}{[n.d.]b}]%
        {RN250}
 \bibinfo{year}{[n.d.]}\natexlab{b}.
\newblock
\newblock
\urldef\tempurl%
\url{http://www.w3.org/Submission/SWRL/}
\showURL{%
\tempurl}


\bibitem[\protect\citeauthoryear{??}{Inv}{[n.d.]}]%
        {InvalidIssue1}
 \bibinfo{year}{[n.d.]}\natexlab{}.
\newblock
\newblock
\urldef\tempurl%
\url{https://github.com/Pryaxis/handbook/issues/3}
\showURL{%
\tempurl}


\bibitem[\protect\citeauthoryear{??}{S1}{[n.d.]}]%
        {S1}
 \bibinfo{year}{[n.d.]}\natexlab{}.
\newblock
\newblock
\urldef\tempurl%
\url{https://github.com/novus-package-manager/novus/issues/3}
\showURL{%
\tempurl}


\bibitem[\protect\citeauthoryear{??}{S2}{[n.d.]}]%
        {S2}
 \bibinfo{year}{[n.d.]}\natexlab{}.
\newblock
\newblock
\urldef\tempurl%
\url{https://github.com/biddyweb/yes-cart/issues/33}
\showURL{%
\tempurl}


\bibitem[\protect\citeauthoryear{??}{S3}{[n.d.]}]%
        {S3}
 \bibinfo{year}{[n.d.]}\natexlab{}.
\newblock
\newblock
\urldef\tempurl%
\url{https://github.com/CircuitVerse/Interactive-Book/issues/80}
\showURL{%
\tempurl}


\bibitem[\protect\citeauthoryear{??}{S4}{[n.d.]}]%
        {S4}
 \bibinfo{year}{[n.d.]}\natexlab{}.
\newblock
\newblock
\urldef\tempurl%
\url{https://github.com/mpdf/mpdf/issues/15}
\showURL{%
\tempurl}


\bibitem[\protect\citeauthoryear{??}{S5}{[n.d.]}]%
        {S5}
 \bibinfo{year}{[n.d.]}\natexlab{}.
\newblock
\newblock
\urldef\tempurl%
\url{https://github.com/pkalogiros/AudioMass/issues/1}
\showURL{%
\tempurl}


\bibitem[\protect\citeauthoryear{??}{S6}{[n.d.]}]%
        {S6}
 \bibinfo{year}{[n.d.]}\natexlab{}.
\newblock
\newblock
\urldef\tempurl%
\url{https://github.com/minio/minio/issues/12143}
\showURL{%
\tempurl}


\bibitem[\protect\citeauthoryear{??}{S7}{[n.d.]}]%
        {S7}
 \bibinfo{year}{[n.d.]}\natexlab{}.
\newblock
\newblock
\urldef\tempurl%
\url{https://github.com/wger-project/wger/issues/266}
\showURL{%
\tempurl}


\bibitem[\protect\citeauthoryear{??}{S9}{[n.d.]}]%
        {S9}
 \bibinfo{year}{[n.d.]}\natexlab{}.
\newblock
\newblock
\urldef\tempurl%
\url{https://github.com/tranleduy2000/javaide/issues/236}
\showURL{%
\tempurl}


\bibitem[\protect\citeauthoryear{??}{S10}{[n.d.]}]%
        {S10}
 \bibinfo{year}{[n.d.]}\natexlab{}.
\newblock
\newblock
\urldef\tempurl%
\url{https://github.com/flyingsaucerproject/flyingsaucer/pull/123}
\showURL{%
\tempurl}


\bibitem[\protect\citeauthoryear{??}{S11}{[n.d.]}]%
        {S11}
 \bibinfo{year}{[n.d.]}\natexlab{}.
\newblock
\newblock
\urldef\tempurl%
\url{https://github.com/click-llc/click-integration-django/issues/1}
\showURL{%
\tempurl}


\bibitem[\protect\citeauthoryear{??}{S12}{[n.d.]}]%
        {S12}
 \bibinfo{year}{[n.d.]}\natexlab{}.
\newblock
\newblock
\urldef\tempurl%
\url{https://github.com/twbs/bootstrap/issues/5632}
\showURL{%
\tempurl}


\bibitem[\protect\citeauthoryear{??}{S13}{[n.d.]}]%
        {S13}
 \bibinfo{year}{[n.d.]}\natexlab{}.
\newblock
\newblock
\urldef\tempurl%
\url{https://github.com/NetHack/NetHack/issues/359}
\showURL{%
\tempurl}


\bibitem[\protect\citeauthoryear{??}{S14}{[n.d.]}]%
        {S14}
 \bibinfo{year}{[n.d.]}\natexlab{}.
\newblock
\newblock
\urldef\tempurl%
\url{https://github.com/EasyEngine/easyengine/issues/488}
\showURL{%
\tempurl}


\bibitem[\protect\citeauthoryear{??}{S15}{[n.d.]}]%
        {S15}
 \bibinfo{year}{[n.d.]}\natexlab{}.
\newblock
\newblock
\urldef\tempurl%
\url{https://github.com/katzwebservices/Contact-Form-7-Newsletter/issues/79}
\showURL{%
\tempurl}


\bibitem[\protect\citeauthoryear{??}{Rep}{[n.d.]}]%
        {Repositories}
 \bibinfo{year}{[n.d.]}\natexlab{}.
\newblock
\newblock
\urldef\tempurl%
\url{https://docs.github.com/en/rest/repos}
\showURL{%
\tempurl}


\bibitem[\protect\citeauthoryear{??}{Cop}{[n.d.]}]%
        {CopyrightAct1968}
 \bibinfo{year}{[n.d.]}\natexlab{}.
\newblock
\newblock
\urldef\tempurl%
\url{https://www.legislation.gov.au/Details/C2017C00180}
\showURL{%
\tempurl}


\bibitem[\protect\citeauthoryear{??}{AC2}{[n.d.]}]%
        {AC2}
 \bibinfo{year}{[n.d.]}\natexlab{}.
\newblock
\newblock
\urldef\tempurl%
\url{https://github.com/manuel-freire/ac2}
\showURL{%
\tempurl}


\bibitem[\protect\citeauthoryear{??}{Add}{[n.d.]}]%
        {AddingLicenseToRepository}
 \bibinfo{year}{[n.d.]}\natexlab{}.
\newblock
\newblock
\urldef\tempurl%
\url{https://docs.github.com/en/communities/setting-up-your-project-for-healthy-contributions/adding-a-license-to-a-repository}
\showURL{%
\tempurl}


\bibitem[\protect\citeauthoryear{??}{Lic}{[n.d.]}]%
        {LicensingRepository}
 \bibinfo{year}{[n.d.]}\natexlab{}.
\newblock
\newblock
\urldef\tempurl%
\url{https://docs.github.com/en/repositories/managing-your-repositorys-settings-and-features/customizing-your-repository/licensing-a-repository}
\showURL{%
\tempurl}


\bibitem[\protect\citeauthoryear{??}{Git}{[n.d.]}]%
        {GitHubAPIPython}
 \bibinfo{year}{[n.d.]}\natexlab{}.
\newblock
\newblock
\urldef\tempurl%
\url{https://github.com/PyGithub/PyGithub}
\showURL{%
\tempurl}


\bibitem[\protect\citeauthoryear{??}{Sel}{[n.d.]}]%
        {SelfPromotionFalsePositive1}
 \bibinfo{year}{[n.d.]}\natexlab{}.
\newblock
\newblock
\urldef\tempurl%
\url{https://github.com/Anarios/return-youtube-dislike/issues/401}
\showURL{%
\tempurl}


\bibitem[\protect\citeauthoryear{??}{dis}{[n.d.]}]%
        {disableissue}
 \bibinfo{year}{[n.d.]}\natexlab{}.
\newblock
\newblock
\urldef\tempurl%
\url{https://docs.github.com/en/repositories/managing-your-repositorys-settings-and-features/enabling-features-for-your-repository/disabling-issues}
\showURL{%
\tempurl}


\bibitem[\protect\citeauthoryear{??}{noi}{[n.d.]}]%
        {noissuetagsample}
 \bibinfo{year}{[n.d.]}\natexlab{}.
\newblock
\newblock
\urldef\tempurl%
\url{https://github.com/rydercalmdown/package_theft_preventor}
\showURL{%
\tempurl}


\bibitem[\protect\citeauthoryear{??}{Eto}{[n.d.]}]%
        {EtorGitHub}
 \bibinfo{year}{[n.d.]}\natexlab{}.
\newblock
\newblock
\urldef\tempurl%
\url{https://github.com/EtorChecker/Etor}
\showURL{%
\tempurl}


\bibitem[\protect\citeauthoryear{??}{url}{[n.d.]a}]%
        {url28}
 \bibinfo{year}{[n.d.]}\natexlab{a}.
\newblock \bibinfo{booktitle}{\emph{ailab}}.
\newblock
\urldef\tempurl%
\url{https://github.com/bilibili/ailab}
\showURL{%
\tempurl}


\bibitem[\protect\citeauthoryear{??}{url}{[n.d.]b}]%
        {url27}
 \bibinfo{year}{[n.d.]}\natexlab{b}.
\newblock \bibinfo{booktitle}{\emph{Are we correctly handling console.Console
  in node objectKeys(console)?}}
\newblock
\urldef\tempurl%
\url{https://github.com/sindresorhus/ts-extras/issues/50}
\showURL{%
\tempurl}


\bibitem[\protect\citeauthoryear{??}{url}{[n.d.]c}]%
        {url25}
 \bibinfo{year}{[n.d.]}\natexlab{c}.
\newblock \bibinfo{booktitle}{\emph{CUDA vs Naive Speedup?}}
\newblock
\urldef\tempurl%
\url{https://github.com/d-li14/involution/issues/1}
\showURL{%
\tempurl}


\bibitem[\protect\citeauthoryear{??}{url}{[n.d.]d}]%
        {url30}
 \bibinfo{year}{[n.d.]}\natexlab{d}.
\newblock \bibinfo{booktitle}{\emph{DogeBot2}}.
\newblock
\urldef\tempurl%
\url{https://github.com/DGXeon/DogeBot2}
\showURL{%
\tempurl}


\bibitem[\protect\citeauthoryear{??}{url}{[n.d.]e}]%
        {url26}
 \bibinfo{year}{[n.d.]}\natexlab{e}.
\newblock \bibinfo{booktitle}{\emph{Squeeze tooltip in the sections panel}}.
\newblock
\urldef\tempurl%
\url{https://github.com/livebook-dev/livebook/pull/536}
\showURL{%
\tempurl}


\bibitem[\protect\citeauthoryear{??}{url}{[n.d.]f}]%
        {url29}
 \bibinfo{year}{[n.d.]}\natexlab{f}.
\newblock \bibinfo{booktitle}{\emph{VIP}}.
\newblock
\urldef\tempurl%
\url{https://github.com/Oreomeow/VIP}
\showURL{%
\tempurl}


\bibitem[\protect\citeauthoryear{??}{RN2}{[n.d.]c}]%
        {RN295}
 \bibinfo{year}{[n.d.]}\natexlab{c}.
\newblock \showarticletitle{What is Plagiarism?}
\newblock  (\bibinfo{year}{[n.\,d.]}).
\newblock
\urldef\tempurl%
\url{https://www.plagiarism.org/article/what-is-plagiarism}
\showURL{%
\tempurl}


\bibitem[\protect\citeauthoryear{??}{lin}{2021}]%
        {linuxhypo}
 \bibinfo{year}{2021}\natexlab{}.
\newblock , \bibinfo{numpages}{Report on University of Minnesota
  Breach-of-Trust Incident}~pages.
\newblock
\urldef\tempurl%
\url{https://lwn.net/ml/linux-kernel/202105051005.49BFABCE@keescook/}
\showURL{%
\tempurl}


\bibitem[\protect\citeauthoryear{Andrews and Pradhan}{Andrews and
  Pradhan}{2001}]%
        {LimitsOfPolicy}
\bibfield{author}{\bibinfo{person}{Anneliese~Amschler Andrews} {and}
  \bibinfo{person}{Arundeep S \%J Empirical Software~Engineering Pradhan}.}
  \bibinfo{year}{2001}\natexlab{}.
\newblock \showarticletitle{Ethical issues in empirical software engineering:
  the limits of policy}.
\newblock  \bibinfo{volume}{6}, \bibinfo{number}{2} (\bibinfo{year}{2001}),
  \bibinfo{pages}{105--110}.
\newblock
\showISSN{1573-7616}


\bibitem[\protect\citeauthoryear{Antoniou and Harmelen}{Antoniou and
  Harmelen}{2004}]%
        {antoniou2004web}
\bibfield{author}{\bibinfo{person}{Grigoris Antoniou} {and}
  \bibinfo{person}{Frank~van Harmelen}.} \bibinfo{year}{2004}\natexlab{}.
\newblock \showarticletitle{Web ontology language: Owl}.
\newblock In \bibinfo{booktitle}{\emph{Handbook on ontologies}}.
  \bibinfo{publisher}{Springer}, \bibinfo{pages}{67--92}.
\newblock


\bibitem[\protect\citeauthoryear{Badampudi}{Badampudi}{[n.d.]}]%
        {ReportingEthics}
\bibfield{author}{\bibinfo{person}{Deepika Badampudi}.}
  \bibinfo{year}{[n.d.]}\natexlab{}.
\newblock \showarticletitle{Reporting ethics considerations in software
  engineering publications}. In \bibinfo{booktitle}{\emph{2017 ACM/IEEE
  International Symposium on Empirical Software Engineering and Measurement
  (ESEM)}}. \bibinfo{publisher}{IEEE}, \bibinfo{pages}{205--210}.
\newblock
\showISBNx{1509040390}


\bibitem[\protect\citeauthoryear{Baltes and Diehl}{Baltes and Diehl}{2016}]%
        {baltes2016worse}
\bibfield{author}{\bibinfo{person}{Sebastian Baltes} {and}
  \bibinfo{person}{Stephan Diehl}.} \bibinfo{year}{2016}\natexlab{}.
\newblock \showarticletitle{Worse than spam: Issues in sampling software
  developers}. In \bibinfo{booktitle}{\emph{Proceedings of the 10th ACM/IEEE
  international symposium on empirical software engineering and measurement}}.
  \bibinfo{pages}{1--6}.
\newblock


\bibitem[\protect\citeauthoryear{Baltes and Diehl}{Baltes and Diehl}{2019}]%
        {baltes2019usage}
\bibfield{author}{\bibinfo{person}{Sebastian Baltes} {and}
  \bibinfo{person}{Stephan Diehl}.} \bibinfo{year}{2019}\natexlab{}.
\newblock \showarticletitle{Usage and attribution of Stack Overflow code
  snippets in GitHub projects}.
\newblock \bibinfo{journal}{\emph{Empirical Software Engineering}}
  \bibinfo{volume}{24}, \bibinfo{number}{3} (\bibinfo{year}{2019}),
  \bibinfo{pages}{1259--1295}.
\newblock


\bibitem[\protect\citeauthoryear{Baltes, Kiefer, and Diehl}{Baltes
  et~al\mbox{.}}{2017}]%
        {baltes2017attribution}
\bibfield{author}{\bibinfo{person}{Sebastian Baltes}, \bibinfo{person}{Richard
  Kiefer}, {and} \bibinfo{person}{Stephan Diehl}.}
  \bibinfo{year}{2017}\natexlab{}.
\newblock \showarticletitle{Attribution required: Stack overflow code snippets
  in GitHub projects}. In \bibinfo{booktitle}{\emph{2017 IEEE/ACM 39th
  International Conference on Software Engineering Companion (ICSE-C)}}. IEEE,
  \bibinfo{pages}{161--163}.
\newblock


\bibitem[\protect\citeauthoryear{Beimel and Peleg}{Beimel and Peleg}{2011}]%
        {beimel2011using}
\bibfield{author}{\bibinfo{person}{Dizza Beimel} {and} \bibinfo{person}{Mor
  Peleg}.} \bibinfo{year}{2011}\natexlab{}.
\newblock \showarticletitle{Using OWL and SWRL to represent and reason with
  situation-based access control policies}.
\newblock \bibinfo{journal}{\emph{Data \& Knowledge Engineering}}
  \bibinfo{volume}{70}, \bibinfo{number}{6} (\bibinfo{year}{2011}),
  \bibinfo{pages}{596--615}.
\newblock


\bibitem[\protect\citeauthoryear{Bergerson}{Bergerson}{2000}]%
        {bergerson2000commerce}
\bibfield{author}{\bibinfo{person}{Stephen~R Bergerson}.}
  \bibinfo{year}{2000}\natexlab{}.
\newblock \showarticletitle{E-commerce Privacy and the Black Hole of
  Cyberspace}.
\newblock \bibinfo{journal}{\emph{Wm. Mitchell L. Rev.}}  \bibinfo{volume}{27}
  (\bibinfo{year}{2000}), \bibinfo{pages}{1527}.
\newblock


\bibitem[\protect\citeauthoryear{Boussi~Rahmouni, Solomonides, Casassa~Mont,
  and Shiu}{Boussi~Rahmouni et~al\mbox{.}}{2009}]%
        {boussi2009modelling}
\bibfield{author}{\bibinfo{person}{Hanene Boussi~Rahmouni},
  \bibinfo{person}{Tony Solomonides}, \bibinfo{person}{Marco Casassa~Mont},
  {and} \bibinfo{person}{Simon Shiu}.} \bibinfo{year}{2009}\natexlab{}.
\newblock \showarticletitle{Modelling and enforcing privacy for medical data
  disclosure across Europe}.
\newblock In \bibinfo{booktitle}{\emph{Medical Informatics in a United and
  Healthy Europe}}. \bibinfo{publisher}{IOS Press}, \bibinfo{pages}{695--699}.
\newblock


\bibitem[\protect\citeauthoryear{Cenite, Detenber, Koh, Lim, Soon, and
  Society}{Cenite et~al\mbox{.}}{2009}]%
        {rightThingOnline}
\bibfield{author}{\bibinfo{person}{Mark Cenite}, \bibinfo{person}{Benjamin~H
  Detenber}, \bibinfo{person}{Andy~WK Koh}, \bibinfo{person}{Alvin~LH Lim},
  \bibinfo{person}{Ng~Ee \%J New~Media Soon}, {and} \bibinfo{person}{Society}.}
  \bibinfo{year}{2009}\natexlab{}.
\newblock \showarticletitle{Doing the right thing online: a survey of bloggers'
  ethical beliefs and practices}.
\newblock  \bibinfo{volume}{11}, \bibinfo{number}{4} (\bibinfo{year}{2009}),
  \bibinfo{pages}{575--597}.
\newblock
\showISSN{1461-4448}


\bibitem[\protect\citeauthoryear{Colquitt}{Colquitt}{2001}]%
        {colquitt2001dimensionality}
\bibfield{author}{\bibinfo{person}{Jason~A Colquitt}.}
  \bibinfo{year}{2001}\natexlab{}.
\newblock \showarticletitle{On the dimensionality of organizational justice: a
  construct validation of a measure.}
\newblock \bibinfo{journal}{\emph{Journal of applied psychology}}
  \bibinfo{volume}{86}, \bibinfo{number}{3} (\bibinfo{year}{2001}),
  \bibinfo{pages}{386}.
\newblock


\bibitem[\protect\citeauthoryear{Cruzes and Dyba}{Cruzes and Dyba}{2011}]%
        {cruzes2011recommended}
\bibfield{author}{\bibinfo{person}{Daniela~S Cruzes} {and}
  \bibinfo{person}{Tore Dyba}.} \bibinfo{year}{2011}\natexlab{}.
\newblock \showarticletitle{Recommended steps for thematic synthesis in
  software engineering}. In \bibinfo{booktitle}{\emph{2011 international
  symposium on empirical software engineering and measurement}}. IEEE,
  \bibinfo{pages}{275--284}.
\newblock


\bibitem[\protect\citeauthoryear{da~Silva, Louro, Goncalves, Marques, Dias,
  da~Cunha, and Tasinaffo}{da~Silva et~al\mbox{.}}{2021}]%
        {da2021could}
\bibfield{author}{\bibinfo{person}{Daniela~America da Silva},
  \bibinfo{person}{Henrique Duarte~Borges Louro},
  \bibinfo{person}{Gildarcio~Sousa Goncalves}, \bibinfo{person}{Johnny~Cardoso
  Marques}, \bibinfo{person}{Luiz Alberto~Vieira Dias},
  \bibinfo{person}{Adilson~Marques da Cunha}, {and}
  \bibinfo{person}{Paulo~Marcelo Tasinaffo}.} \bibinfo{year}{2021}\natexlab{}.
\newblock \showarticletitle{Could a Conversational AI Identify Offensive
  Language?}
\newblock \bibinfo{journal}{\emph{Information}} \bibinfo{volume}{12},
  \bibinfo{number}{10} (\bibinfo{year}{2021}), \bibinfo{pages}{418}.
\newblock


\bibitem[\protect\citeauthoryear{Eisenbarth, Koschke, and Simon}{Eisenbarth
  et~al\mbox{.}}{2003}]%
        {eisenbarth2003locating}
\bibfield{author}{\bibinfo{person}{Thomas Eisenbarth}, \bibinfo{person}{Rainer
  Koschke}, {and} \bibinfo{person}{Daniel Simon}.}
  \bibinfo{year}{2003}\natexlab{}.
\newblock \showarticletitle{Locating features in source code}.
\newblock \bibinfo{journal}{\emph{IEEE Transactions on software engineering}}
  \bibinfo{volume}{29}, \bibinfo{number}{3} (\bibinfo{year}{2003}),
  \bibinfo{pages}{210--224}.
\newblock


\bibitem[\protect\citeauthoryear{Friedman, Kahn, Borning, and
  Huldtgren}{Friedman et~al\mbox{.}}{2013}]%
        {ValueSensitive}
\bibfield{author}{\bibinfo{person}{Batya Friedman}, \bibinfo{person}{Peter~H
  Kahn}, \bibinfo{person}{Alan Borning}, {and} \bibinfo{person}{Alina
  Huldtgren}.} \bibinfo{year}{2013}\natexlab{}.
\newblock \bibinfo{booktitle}{\emph{Value sensitive design and information
  systems}}.
\newblock \bibinfo{publisher}{Springer}, \bibinfo{pages}{55--95}.
\newblock


\bibitem[\protect\citeauthoryear{German, Manabe, and Inoue}{German
  et~al\mbox{.}}{2010}]%
        {german2010sentence}
\bibfield{author}{\bibinfo{person}{Daniel~M German}, \bibinfo{person}{Yuki
  Manabe}, {and} \bibinfo{person}{Katsuro Inoue}.}
  \bibinfo{year}{2010}\natexlab{}.
\newblock \showarticletitle{A sentence-matching method for automatic license
  identification of source code files}. In
  \bibinfo{booktitle}{\emph{Proceedings of the IEEE/ACM international
  conference on Automated software engineering}}. \bibinfo{pages}{437--446}.
\newblock


\bibitem[\protect\citeauthoryear{German, Robles, Poo-Caama{\~n}o, Yang, Iida,
  and Inoue}{German et~al\mbox{.}}{2018}]%
        {german2018my}
\bibfield{author}{\bibinfo{person}{Daniel~M German}, \bibinfo{person}{Gregorio
  Robles}, \bibinfo{person}{Germ{\'a}n Poo-Caama{\~n}o}, \bibinfo{person}{Xin
  Yang}, \bibinfo{person}{Hajimu Iida}, {and} \bibinfo{person}{Katsuro Inoue}.}
  \bibinfo{year}{2018}\natexlab{}.
\newblock \showarticletitle{"Was My Contribution Fairly Reviewed?" A Framework
  to Study the Perception of Fairness in Modern Code Reviews}. In
  \bibinfo{booktitle}{\emph{2018 IEEE/ACM 40th International Conference on
  Software Engineering (ICSE)}}. IEEE, \bibinfo{pages}{523--534}.
\newblock


\bibitem[\protect\citeauthoryear{Gold and Krinke}{Gold and Krinke}{[n.d.]}]%
        {EthicalMining}
\bibfield{author}{\bibinfo{person}{Nicolas~E Gold} {and} \bibinfo{person}{Jens
  Krinke}.} \bibinfo{year}{[n.d.]}\natexlab{}.
\newblock \showarticletitle{Ethical Mining: A Case Study on MSR Mining
  Challenges}. In \bibinfo{booktitle}{\emph{Proceedings of the 17th
  International Conference on Mining Software Repositories}}.
  \bibinfo{pages}{265--276}.
\newblock


\bibitem[\protect\citeauthoryear{Golubev, Eliseeva, Povarov, and
  Bryksin}{Golubev et~al\mbox{.}}{2020}]%
        {golubev2020study}
\bibfield{author}{\bibinfo{person}{Yaroslav Golubev}, \bibinfo{person}{Maria
  Eliseeva}, \bibinfo{person}{Nikita Povarov}, {and} \bibinfo{person}{Timofey
  Bryksin}.} \bibinfo{year}{2020}\natexlab{}.
\newblock \showarticletitle{A study of potential code borrowing and license
  violations in java projects on github}. In
  \bibinfo{booktitle}{\emph{Proceedings of the 17th International Conference on
  Mining Software Repositories}}. \bibinfo{pages}{54--64}.
\newblock


\bibitem[\protect\citeauthoryear{Grodzinsky, Miller, and Wolf}{Grodzinsky
  et~al\mbox{.}}{2003}]%
        {EthicalIssuesInOSS}
\bibfield{author}{\bibinfo{person}{Frances~S Grodzinsky},
  \bibinfo{person}{Keith Miller}, {and} \bibinfo{person}{Marty~J Wolf}.}
  \bibinfo{year}{2003}\natexlab{}.
\newblock \showarticletitle{Ethical issues in open source software}.
\newblock \bibinfo{journal}{\emph{Journal of Information, Communication and
  Ethics in Society}} (\bibinfo{year}{2003}).
\newblock


\bibitem[\protect\citeauthoryear{Hsi and Potts}{Hsi and Potts}{2000}]%
        {hsi2000studying}
\bibfield{author}{\bibinfo{person}{Idris Hsi} {and} \bibinfo{person}{Colin
  Potts}.} \bibinfo{year}{2000}\natexlab{}.
\newblock \showarticletitle{Studying the Evolution and Enhancement of Software
  Features.}. In \bibinfo{booktitle}{\emph{icsm}}. \bibinfo{pages}{143}.
\newblock


\bibitem[\protect\citeauthoryear{Huq, Sadiq, and Sakib}{Huq
  et~al\mbox{.}}{2019}]%
        {huq2019understanding}
\bibfield{author}{\bibinfo{person}{Syed~Fatiul Huq}, \bibinfo{person}{Ali~Zafar
  Sadiq}, {and} \bibinfo{person}{Kazi Sakib}.} \bibinfo{year}{2019}\natexlab{}.
\newblock \showarticletitle{Understanding the effect of developer sentiment on
  fix-inducing changes: An exploratory study on github pull requests}. In
  \bibinfo{booktitle}{\emph{2019 26th Asia-Pacific Software Engineering
  Conference (APSEC)}}. IEEE, \bibinfo{pages}{514--521}.
\newblock


\bibitem[\protect\citeauthoryear{Imtiaz, Middleton, Chakraborty, Robson, Bai,
  and Murphy-Hill}{Imtiaz et~al\mbox{.}}{[n.d.]}]%
        {InvestigatingBiasGitHub}
\bibfield{author}{\bibinfo{person}{Nasif Imtiaz}, \bibinfo{person}{Justin
  Middleton}, \bibinfo{person}{Joymallya Chakraborty}, \bibinfo{person}{Neill
  Robson}, \bibinfo{person}{Gina Bai}, {and} \bibinfo{person}{Emerson
  Murphy-Hill}.} \bibinfo{year}{[n.d.]}\natexlab{}.
\newblock \showarticletitle{Investigating the effects of gender bias on
  GitHub}. In \bibinfo{booktitle}{\emph{2019 IEEE/ACM 41st International
  Conference on Software Engineering (ICSE)}}. \bibinfo{publisher}{IEEE},
  \bibinfo{pages}{700--711}.
\newblock
\showISBNx{1728108691}


\bibitem[\protect\citeauthoryear{Kapitsaki, Kramer, and Tselikas}{Kapitsaki
  et~al\mbox{.}}{2017}]%
        {kapitsaki2017automating}
\bibfield{author}{\bibinfo{person}{Georgia~M Kapitsaki},
  \bibinfo{person}{Frederik Kramer}, {and} \bibinfo{person}{Nikolaos~D
  Tselikas}.} \bibinfo{year}{2017}\natexlab{}.
\newblock \showarticletitle{Automating the license compatibility process in
  open source software with SPDX}.
\newblock \bibinfo{journal}{\emph{Journal of systems and software}}
  \bibinfo{volume}{131} (\bibinfo{year}{2017}), \bibinfo{pages}{386--401}.
\newblock


\bibitem[\protect\citeauthoryear{Kapitsaki, Tselikas, and Foukarakis}{Kapitsaki
  et~al\mbox{.}}{2015}]%
        {kapitsaki2015insight}
\bibfield{author}{\bibinfo{person}{Georgia~M Kapitsaki},
  \bibinfo{person}{Nikolaos~D Tselikas}, {and} \bibinfo{person}{Ioannis~E
  Foukarakis}.} \bibinfo{year}{2015}\natexlab{}.
\newblock \showarticletitle{An insight into license tools for open source
  software systems}.
\newblock \bibinfo{journal}{\emph{Journal of Systems and Software}}
  \bibinfo{volume}{102} (\bibinfo{year}{2015}), \bibinfo{pages}{72--87}.
\newblock


\bibitem[\protect\citeauthoryear{Kayes, Rahayu, Dillon, and Chang}{Kayes
  et~al\mbox{.}}{2018}]%
        {kayes2018accessing}
\bibfield{author}{\bibinfo{person}{ASM Kayes}, \bibinfo{person}{Wenny Rahayu},
  \bibinfo{person}{Tharam Dillon}, {and} \bibinfo{person}{Elizabeth Chang}.}
  \bibinfo{year}{2018}\natexlab{}.
\newblock \showarticletitle{Accessing data from multiple sources through
  context-aware access control}. In \bibinfo{booktitle}{\emph{2018 17th IEEE
  International Conference On Trust, Security And Privacy In Computing And
  Communications/12th IEEE International Conference On Big Data Science And
  Engineering (TrustCom/BigDataSE)}}. IEEE, \bibinfo{pages}{551--559}.
\newblock


\bibitem[\protect\citeauthoryear{Kocsis and de~Vreede}{Kocsis and
  de~Vreede}{2016}]%
        {Crowdsourcing}
\bibfield{author}{\bibinfo{person}{David Kocsis} {and}
  \bibinfo{person}{Gert-Jan de Vreede}.} \bibinfo{year}{2016}\natexlab{}.
\newblock \showarticletitle{Towards a taxonomy of ethical considerations in
  crowdsourcing}.
\newblock  (\bibinfo{year}{2016}).
\newblock


\bibitem[\protect\citeauthoryear{Lerner and Tirole}{Lerner and Tirole}{2005}]%
        {lerner2005scope}
\bibfield{author}{\bibinfo{person}{Josh Lerner} {and} \bibinfo{person}{Jean
  Tirole}.} \bibinfo{year}{2005}\natexlab{}.
\newblock \showarticletitle{The scope of open source licensing}.
\newblock \bibinfo{journal}{\emph{Journal of Law, Economics, and Organization}}
  \bibinfo{volume}{21}, \bibinfo{number}{1} (\bibinfo{year}{2005}),
  \bibinfo{pages}{20--56}.
\newblock


\bibitem[\protect\citeauthoryear{McDonnell, Ray, and Kim}{McDonnell
  et~al\mbox{.}}{2013}]%
        {mcdonnell2013empirical}
\bibfield{author}{\bibinfo{person}{Tyler McDonnell}, \bibinfo{person}{Baishakhi
  Ray}, {and} \bibinfo{person}{Miryung Kim}.} \bibinfo{year}{2013}\natexlab{}.
\newblock \showarticletitle{An empirical study of api stability and adoption in
  the android ecosystem}. In \bibinfo{booktitle}{\emph{2013 IEEE International
  Conference on Software Maintenance}}. IEEE, \bibinfo{pages}{70--79}.
\newblock


\bibitem[\protect\citeauthoryear{McGuinness, Van~Harmelen,
  et~al\mbox{.}}{McGuinness et~al\mbox{.}}{2004}]%
        {mcguinness2004owl}
\bibfield{author}{\bibinfo{person}{Deborah~L McGuinness},
  \bibinfo{person}{Frank Van~Harmelen}, {et~al\mbox{.}}}
  \bibinfo{year}{2004}\natexlab{}.
\newblock \showarticletitle{OWL web ontology language overview}.
\newblock \bibinfo{journal}{\emph{W3C recommendation}} \bibinfo{volume}{10},
  \bibinfo{number}{10} (\bibinfo{year}{2004}), \bibinfo{pages}{2004}.
\newblock


\bibitem[\protect\citeauthoryear{McIlroy, Ali, and Hassan}{McIlroy
  et~al\mbox{.}}{2016}]%
        {mcilroy2016fresh}
\bibfield{author}{\bibinfo{person}{Stuart McIlroy}, \bibinfo{person}{Nasir
  Ali}, {and} \bibinfo{person}{Ahmed~E Hassan}.}
  \bibinfo{year}{2016}\natexlab{}.
\newblock \showarticletitle{Fresh apps: an empirical study of
  frequently-updated mobile apps in the Google play store}.
\newblock \bibinfo{journal}{\emph{Empirical Software Engineering}}
  \bibinfo{volume}{21}, \bibinfo{number}{3} (\bibinfo{year}{2016}),
  \bibinfo{pages}{1346--1370}.
\newblock


\bibitem[\protect\citeauthoryear{McNamara, Smith, and Murphy-Hill}{McNamara
  et~al\mbox{.}}{[n.d.]}]%
        {ACMChanges}
\bibfield{author}{\bibinfo{person}{Andrew McNamara}, \bibinfo{person}{Justin
  Smith}, {and} \bibinfo{person}{Emerson Murphy-Hill}.}
  \bibinfo{year}{[n.d.]}\natexlab{}.
\newblock \showarticletitle{Does ACM’s code of ethics change ethical decision
  making in software development?}. In \bibinfo{booktitle}{\emph{Proceedings of
  the 2018 26th ACM joint meeting on european software engineering conference
  and symposium on the foundations of software engineering}}.
  \bibinfo{pages}{729--733}.
\newblock


\bibitem[\protect\citeauthoryear{Mittelstadt}{Mittelstadt}{2019}]%
        {aiaethical}
\bibfield{author}{\bibinfo{person}{Brent Mittelstadt}.}
  \bibinfo{year}{2019}\natexlab{}.
\newblock \showarticletitle{Principles alone cannot guarantee ethical AI}.
\newblock \bibinfo{journal}{\emph{Nature Machine Intelligence}}
  \bibinfo{volume}{1} (\bibinfo{date}{11} \bibinfo{year}{2019}).
\newblock
\urldef\tempurl%
\url{https://doi.org/10.1038/s42256-019-0114-4}
\showDOI{\tempurl}


\bibitem[\protect\citeauthoryear{Mondal, Silva, and Benevenuto}{Mondal
  et~al\mbox{.}}{2017}]%
        {mondal2017measurement}
\bibfield{author}{\bibinfo{person}{Mainack Mondal},
  \bibinfo{person}{Leandro~Ara{\'u}jo Silva}, {and}
  \bibinfo{person}{Fabr{\'\i}cio Benevenuto}.} \bibinfo{year}{2017}\natexlab{}.
\newblock \showarticletitle{A measurement study of hate speech in social
  media}. In \bibinfo{booktitle}{\emph{Proceedings of the 28th ACM conference
  on hypertext and social media}}. \bibinfo{pages}{85--94}.
\newblock


\bibitem[\protect\citeauthoryear{Musen}{Musen}{2015}]%
        {musen2015protege}
\bibfield{author}{\bibinfo{person}{Mark~A Musen}.}
  \bibinfo{year}{2015}\natexlab{}.
\newblock \showarticletitle{The prot{\'e}g{\'e} project: a look back and a look
  forward}.
\newblock \bibinfo{journal}{\emph{AI matters}} \bibinfo{volume}{1},
  \bibinfo{number}{4} (\bibinfo{year}{2015}), \bibinfo{pages}{4--12}.
\newblock


\bibitem[\protect\citeauthoryear{Nyman and Mikkonen}{Nyman and
  Mikkonen}{2011}]%
        {nyman2011fork}
\bibfield{author}{\bibinfo{person}{Linus Nyman} {and} \bibinfo{person}{Tommi
  Mikkonen}.} \bibinfo{year}{2011}\natexlab{}.
\newblock \showarticletitle{To fork or not to fork: Fork motivations in
  SourceForge projects}.
\newblock \bibinfo{journal}{\emph{International Journal of Open Source Software
  and Processes (IJOSSP)}} \bibinfo{volume}{3}, \bibinfo{number}{3}
  (\bibinfo{year}{2011}), \bibinfo{pages}{1--9}.
\newblock


\bibitem[\protect\citeauthoryear{Oezbek et~al\mbox{.}}{Oezbek
  et~al\mbox{.}}{2008}]%
        {ResearchEthicsOpenSource}
\bibfield{author}{\bibinfo{person}{Christopher Oezbek} {et~al\mbox{.}}}
  \bibinfo{year}{2008}\natexlab{}.
\newblock \showarticletitle{Research ethics for studying Open Source projects}.
\newblock \bibinfo{journal}{\emph{4th Research Room FOSDEM: Libre software
  communities meet research community}} (\bibinfo{year}{2008}).
\newblock


\bibitem[\protect\citeauthoryear{Pfeiffer}{Pfeiffer}{2020}]%
        {pfeiffer2020constitutes}
\bibfield{author}{\bibinfo{person}{Rolf-Helge Pfeiffer}.}
  \bibinfo{year}{2020}\natexlab{}.
\newblock \showarticletitle{What constitutes software? An empirical,
  descriptive study of artifacts}. In \bibinfo{booktitle}{\emph{Proceedings of
  the 17th International Conference on Mining Software Repositories}}.
  \bibinfo{pages}{481--491}.
\newblock


\bibitem[\protect\citeauthoryear{Singer and Vinson}{Singer and Vinson}{2002}]%
        {EthicalIssuesEmpiricalStudiesSE}
\bibfield{author}{\bibinfo{person}{Janice Singer} {and} \bibinfo{person}{Norman
  G. \%J IEEE Transactions on Software~Engineering Vinson}.}
  \bibinfo{year}{2002}\natexlab{}.
\newblock \showarticletitle{Ethical issues in empirical studies of software
  engineering}.
\newblock  \bibinfo{volume}{28}, \bibinfo{number}{12} (\bibinfo{year}{2002}),
  \bibinfo{pages}{1171--1180}.
\newblock
\showISSN{0098-5589}


\bibitem[\protect\citeauthoryear{Terrell, Kofink, Middleton, Rainear,
  Murphy-Hill, and Parnin}{Terrell et~al\mbox{.}}{2016}]%
        {terrell2016gender}
\bibfield{author}{\bibinfo{person}{Josh Terrell}, \bibinfo{person}{Andrew
  Kofink}, \bibinfo{person}{Justin Middleton}, \bibinfo{person}{Clarissa
  Rainear}, \bibinfo{person}{Emerson~R Murphy-Hill}, {and}
  \bibinfo{person}{Chris Parnin}.} \bibinfo{year}{2016}\natexlab{}.
\newblock \showarticletitle{Gender bias in open source: Pull request acceptance
  of women versus men.}
\newblock \bibinfo{journal}{\emph{PeerJ Prepr.}}  \bibinfo{volume}{4}
  (\bibinfo{year}{2016}), \bibinfo{pages}{e1733}.
\newblock


\bibitem[\protect\citeauthoryear{Turilli and Floridi}{Turilli and
  Floridi}{2009}]%
        {InformationTransparency}
\bibfield{author}{\bibinfo{person}{Matteo Turilli} {and}
  \bibinfo{person}{Luciano Floridi}.} \bibinfo{year}{2009}\natexlab{}.
\newblock \showarticletitle{The ethics of information transparency}.
\newblock \bibinfo{journal}{\emph{Ethics and Information Technology}}
  \bibinfo{volume}{11}, \bibinfo{number}{2} (\bibinfo{year}{2009}),
  \bibinfo{pages}{105--112}.
\newblock


\bibitem[\protect\citeauthoryear{Vendome, Linares-V{\'a}squez, Bavota,
  Di~Penta, German, and Poshyvanyk}{Vendome et~al\mbox{.}}{2017}]%
        {vendome2017machine}
\bibfield{author}{\bibinfo{person}{Christopher Vendome}, \bibinfo{person}{Mario
  Linares-V{\'a}squez}, \bibinfo{person}{Gabriele Bavota},
  \bibinfo{person}{Massimiliano Di~Penta}, \bibinfo{person}{Daniel German},
  {and} \bibinfo{person}{Denys Poshyvanyk}.} \bibinfo{year}{2017}\natexlab{}.
\newblock \showarticletitle{Machine learning-based detection of open source
  license exceptions}. In \bibinfo{booktitle}{\emph{2017 IEEE/ACM 39th
  International Conference on Software Engineering (ICSE)}}. IEEE,
  \bibinfo{pages}{118--129}.
\newblock


\bibitem[\protect\citeauthoryear{Vendome, Linares-Vásquez, Bavota, Di~Penta,
  German, and Poshyvanyk}{Vendome et~al\mbox{.}}{[n.d.]}]%
        {LicenseUsageChangesJava}
\bibfield{author}{\bibinfo{person}{Christopher Vendome}, \bibinfo{person}{Mario
  Linares-Vásquez}, \bibinfo{person}{Gabriele Bavota},
  \bibinfo{person}{Massimiliano Di~Penta}, \bibinfo{person}{Daniel German},
  {and} \bibinfo{person}{Denys Poshyvanyk}.} \bibinfo{year}{[n.d.]}\natexlab{}.
\newblock \showarticletitle{License usage and changes: a large-scale study of
  java projects on github}. In \bibinfo{booktitle}{\emph{2015 IEEE 23rd
  International Conference on Program Comprehension}}.
  \bibinfo{publisher}{IEEE}, \bibinfo{pages}{218--228}.
\newblock
\showISBNx{1467381594}


\bibitem[\protect\citeauthoryear{Vrande{\v{c}}i{\'c}}{Vrande{\v{c}}i{\'c}}{2009}]%
        {vrandevcic2009ontology}
\bibfield{author}{\bibinfo{person}{Denny Vrande{\v{c}}i{\'c}}.}
  \bibinfo{year}{2009}\natexlab{}.
\newblock \showarticletitle{Ontology evaluation}.
\newblock In \bibinfo{booktitle}{\emph{Handbook on ontologies}}.
  \bibinfo{publisher}{Springer}, \bibinfo{pages}{293--313}.
\newblock


\bibitem[\protect\citeauthoryear{Wu and Lu}{Wu and Lu}{2021}]%
        {wu2021feasibility}
\bibfield{author}{\bibinfo{person}{Qiushi Wu} {and} \bibinfo{person}{Kangjie
  Lu}.} \bibinfo{year}{2021}\natexlab{}.
\newblock \showarticletitle{On the feasibility of stealthily introducing
  vulnerabilities in open-source software via hypocrite commits}. In
  \bibinfo{booktitle}{\emph{Proc. Oakland}}.
\newblock


\bibitem[\protect\citeauthoryear{Xu, Gao, Fan, Liu, Liu, and Ji}{Xu
  et~al\mbox{.}}{2021}]%
        {xu2021lidetector}
\bibfield{author}{\bibinfo{person}{Sihan Xu}, \bibinfo{person}{Ya Gao},
  \bibinfo{person}{Lingling Fan}, \bibinfo{person}{Zheli Liu},
  \bibinfo{person}{Yang Liu}, {and} \bibinfo{person}{Hua Ji}.}
  \bibinfo{year}{2021}\natexlab{}.
\newblock \showarticletitle{LiDetector: License Incompatibility Detection for
  Open Source Software}.
\newblock \bibinfo{journal}{\emph{ACM Transactions on Software Engineering and
  Methodology}} (\bibinfo{year}{2021}).
\newblock


\bibitem[\protect\citeauthoryear{Yang, Martins, Saini, and Lopes}{Yang
  et~al\mbox{.}}{2017}]%
        {yang2017stack}
\bibfield{author}{\bibinfo{person}{Di Yang}, \bibinfo{person}{Pedro Martins},
  \bibinfo{person}{Vaibhav Saini}, {and} \bibinfo{person}{Cristina Lopes}.}
  \bibinfo{year}{2017}\natexlab{}.
\newblock \showarticletitle{Stack overflow in github: any snippets there?}. In
  \bibinfo{booktitle}{\emph{2017 IEEE/ACM 14th International Conference on
  Mining Software Repositories (MSR)}}. IEEE, \bibinfo{pages}{280--290}.
\newblock


\end{thebibliography}

\end{document}